\documentclass[sigconf,nonacm]{acmart}

\AtBeginDocument{%
  }

\setcopyright{acmlicensed}
\copyrightyear{2026}
\acmYear{2026}
\acmDOI{XXXXXXX.XXXXXXX}
\acmConference[ACM ASIACCS 2027]{}{June 03--05,
  2018}{Macau}
\acmISBN{978-1-4503-XXXX-X/2018/06}

\usepackage{natbib}
\usepackage{amsmath,amsfonts}
\usepackage{algorithmic}
\usepackage{graphicx}
\usepackage{textcomp}
\usepackage{xcolor}
\usepackage{booktabs}
\usepackage{url}
\usepackage{subcaption}

\begin{document}

\title{Detecting False Data Injection and Unstable Operation in Smart Grid via System-Aware Graph Boundary Learning}
% Decentralized Smart Grid Control
% Graph Learning of the Stable Operating Region
% Detecting Instability and Unseen FDI Attacks in Smart Grids via Physics-Constrained Graph Learning of the Stable Region
% \correspondingauthor
% \affiliation{%

\author{Emad Efatinasab}
\affiliation{%
  \institution{University of Padua}
  \city{Padova}
  \country{Italy}}
\email{emad.efatinasab@phd.unipd.it}

\author{Denis Donadel}
\affiliation{%
  \institution{Fondazione Bruno Kessler (FBK)}
  \city{Trento }
  \country{Italy}}
\email{ddonadel@fbk.eu}

\author{Mirco Rampazzo}
\affiliation{%
  \institution{University of Padua}
  \city{Padova}
  \country{Italy}}
\email{mirco.rampazzo@unipd.it}

\author{Chuadhry Mujeeb Ahmed}
\affiliation{%
  \institution{Newcastle University}
  \city{Newcastle upon Tyne}
  \country{UK}}
\email{mujeeb.ahmed@newcastle.ac.uk}

% \affiliation{%

% \affiliation{%

% \affiliation{%

% \affiliation{%

% \correspondingauthor
% \affiliation{%

% \renewcommand{\shortauthors}{ et al.}

\begin{abstract}
Cyber-physical power systems increasingly rely on data-driven tools to detect instability and support reliable grid operation. However, reliable stability prediction is difficult when unstable operating configurations are rare, sensitive, or unavailable during model development, since collecting such data safely and at scale is often impractical. At the same time, False Data Injection (FDI) attacks can manipulate reported system parameters to trigger false instability alarms or conceal genuinely unsafe operation, and such threats in Decentral Smart Grid Control (DSGC) systems remain largely unexplored. These two challenges are rarely addressed jointly, leaving a gap between stability prediction and attack detection that this work aims to close.

In this paper, we introduce StarGNN, a graph learning framework that learns the stable operating region exclusively from clean stable configurations and uses a single abnormality score to flag reported configurations that should not be trusted as evidence of safe operation, covering both genuine instability and unseen FDI manipulations. Each configuration is represented as a producer--consumer star graph and processed by a role aware graph neural network, with physics constrained pseudo-negatives generated by perturbing reaction time and price response parameters standing in for the unavailable unstable and attack data. A single threshold, calibrated only on held-out stable data, is used without task or attack specific adjustment. Evaluated on nine unseen FDI scenarios, StarGNN detects between 0.780 and 0.973 of attacks on stable configurations and retains a post attack instability recall between 0.972 and 0.999 on unstable ones, with 0.899 recall against a stronger adaptive attacker, showing that stable-only boundary learning can support both stability prediction and attack detection without access to genuine unstable labels or attack samples during training.
\end{abstract}





\maketitle
% Requires \usepackage{xcolor} in the preamble (for \textcolor{blue}{...})

\section{Introduction}

Cyber-physical power systems are now widely adopted across the electricity sector, where they play a central role in supporting reliable, efficient, and secure grid operation. The operation of cyber-physical power systems relies on interdependencies among the physical power grid, physical communication network, and logical communication network, which support the transmission of operational data and control commands between grid components and control centres~\cite{10634994}. The increasing reliance on cyber-physical technologies has accelerated the evolution of conventional electrical networks toward smart grids.

Maintaining a reliable match between electricity generation and consumption is a key objective of smart grid operation. To achieve this, predictive tools and proactive control measures are employed to anticipate imbalances and limit the likelihood of system disturbances~\cite{EFATINASAB2025101799}. A commonly used approach for aligning electricity consumption with prevailing grid conditions is demand response~\cite{5930335,en13102559}. Within this context, Decentral Smart Grid Control (DSGC)~\cite{schafer2015decentral} provides a local mechanism for dynamic demand response. It uses grid frequency deviations to indicate the prevailing power imbalance: a decrease signals insufficient generation, whereas an increase indicates excess supply. Consequently, electricity prices are determined from local frequency measurements.

However, introducing an electricity price dependent on the frequency causes the economic response of grid participants to interact directly with the physical grid dynamics. Whether the resulting operating point remains locally stable depends on the intensity of this price response, delays in participant adaptation, and the time window used to average frequency measurements~\cite{schafer2015decentral,8587498}. Since the response coefficients and adaptation delays need not be identical across grid participants, identifying whether perturbations around equilibrium will diminish or increase is a challenging task~\cite{8587498}. Unstable grid operation may compromise both power system reliability and economic activity, potentially triggering severe voltage disturbances and, in extreme cases, cascading blackouts across large areas~\cite{efatinasab2024gangrid}.

Previous research has demonstrated that machine learning can effectively support smart grid stability prediction, with consistently high prediction accuracy reported across different model architectures~\cite{9079864,SHI2020115733,AYGUL2024101012}. Despite this strong performance, such supervised approaches share a critical limitation: their practical deployment is constrained by the limited availability of representative instability data. Real operational datasets containing both stable and unstable grid conditions are difficult to obtain because instability events are uncommon, undesirable, and potentially unsafe to reproduce deliberately~\cite{EFATINASAB2025101662}; their collection may also require substantial expertise, resources, and time~\cite{9058539}. Consequently, methods that assume access to unstable examples may be difficult to apply under realistic data constraints, which is the case for most models in the literature.

Beyond this data scarcity, DSGC systems face a second and largely separate challenge: their exposure to cyber threats. Integrating computational and communication capabilities with physical processes can improve system efficiency; however, the use of open, networked communication channels also increases cyber-physical systems' exposure to malicious external interference~\cite{11175202}. Attacks on the cyber infrastructure of electrical grids may interrupt energy services, undermine critical assets, and generate substantial risks for human well-being and environmental security~\cite{ALVAREZALVARADO2024109149}. Among these threats, False Data Injection (FDI) attacks are widely regarded as one of the most critical cyber threats to the integrity of modern power grids~\cite{AOUFI2020102518}, in which adversaries manipulate system data to influence monitoring or control decisions. Although supervised FDI attack detectors can achieve high accuracy when the attack patterns encountered during testing are represented in their training data, detection performance can deteriorate substantially when the characteristics of a new FDI attack differ from those learned during training~\cite{10.1007/978-3-031-47721-8_16}. Despite this plausible attack surface, FDI threats in DSGC systems remain largely unexplored, motivating a detector that addresses natural instability and adversarial manipulation jointly rather than as separate problems.

\paragraph{Contributions} In this paper, we bridge these gaps by proposing \emph{StarGNN}, the first system model-guided graph neural network for unified detection of natural instability and FDI attacks in DSGC systems, trained on stable configuration samples only. StarGNN represents each operating configuration as a producer-consumer star graph with role-aware node processing and assigns a single risk score to how far it departs from stable operation. The model learns the boundary of the stable region without ever observing unstable or attacked configurations during training. This is a key advantage over supervised detectors in the literature, which require instability or attack data that are scarce, unsafe to collect, or, for FDI attacks, largely unexplored in DSGC systems. Our contributions can be summarized as follows:
\begin{itemize}
    \item We propose \emph{StarGNN}, a DSGC model-guided graph neural network that encodes each operating configuration as a producer--consumer star graph with role-aware node processing, assigning a single risk score to departures from stable operation.
 
    \item We introduce a learning setting in which neither unstable configurations nor attack samples are available during training, along with a generation strategy that produces physically plausible synthetic samples by perturbing reaction time and price-response parameters.
 
    \item We formulate a single detector and threshold, calibrated only on held-out stable data, for both natural instability and FDI manipulation, without task or attack-specific adjustment. Moreover, we provide a security evaluation of StarGNN covering nine unseen FDI scenarios and a constrained adaptive attacker with binary alarm feedback.
 
    \item We show that StarGNN reaches 0.989 accuracy and a macro F1-score of 0.97 on stability prediction. Under attack, it maintains a mean attack recall of 0.988 against generic FDI scenarios, while still reaching an attack recall of 0.899 against an adaptive attacker, consistently outperforming established anomaly detection baselines.
    %Code and data are available.\footnote{\url{https://anonymous.4open.science/r/System-Aware-Graph-Boundary-Learning-8E0E}}}
\end{itemize}

\paragraph{Organization} The remainder of this paper is organized as follows. Section~\ref{sec:related} reviews related work. Section~\ref{sec:system_model} and Section~\ref{sec:threat_model} introduce the considered system and threat model, respectively. Section~\ref{sec:fdi} introduces the attacks we employed in this study. StarGNN is illustrated in Section~\ref{sec:methodology} and evaluated in Section~\ref{sec:evaluation}. Finally, Section~\ref{sec:discussion} offers some discussion insights, while Section~\ref{sec:conclusion} concludes this paper with some final remarks.
\section{Related Work}\label{sec:related}

\paragraph{AI-based Stability Prediction.}
Recent advances in artificial intelligence have significantly expanded the capabilities of smart-grid stability assessment and control~\cite{SHI2020115733}.
For instance , Wang et al.~\cite{WANG2025101175} compared six machine learning models such as Random Forest (RF), Extreme Gradient Boosting (XGBoost), Support Vector Machine (SVM), K-Nearest Neighbors (KNN), Logistic Regression (LR), and Categorical Boosting (CatBoost) for smart grid stability classification and prediction. Another study introduced a pruned one-dimensional time-aware CNN for grid stability analysis and energy cost optimization~\cite{AHAKONYE2024101086}.
A Multidirectional LSTM model is proposed by~\cite{9079864} for stability prediction. Numerous studies have investigated smart grid stability using a range of machine learning and artificial intelligence techniques~\cite{onder2023classification,en18133431,ANIS2026102083}. However, most existing approaches rely on supervised learning and assume access to representative samples from both stable and unstable classes, which may limit their practical deployment when instability data are scarce or unavailable in real world situations. Furthermore, to the best of our knowledge, prior work has not represented DSGC operating configurations as graph structured data to explicitly capture the dependencies between the producer and participating consumers.

\paragraph{False Data Injection Attacks.}
FDI attacks compromise measurement integrity and pose a serious threat to supervisory control and data acquisition systems~\cite{7926429}. These concerns have motivated growing interest in developing effective detection mechanisms for FDI attacks. For instance, Zhang et al.~\cite{9144530} proposed a data-driven method for detecting FDI attacks in distribution system using Autoencoders for dimensionality reduction and feature extraction, and a GAN-based framework that identified attacks by learning discrepancies between secure and manipulated measurements. 

Takiddin et al.~\cite{10016905} proposed a graph autoencoder-based detector that learns spatio-temporal power system features from multiple topology realizations, improving generalization to unseen FDI attacks. 
Despite the growing literature on FDI detection in power systems, the corresponding threat has received limited attention in the context of DSGC systems. Previous studies have examined adversarial attacks against AI-based stability predictors~\cite{efatinasab2024gangrid,EFATINASAB2025101662}. Such attacks represent a related form of data integrity manipulation, but they are typically designed to exploit the decision boundary of a particular prediction model. In contrast, the non-adaptive FDI scenarios considered in this work are motivated by power system attack mechanisms and the structure of the DSGC setting. Their construction reflects participant roles, producer--consumer interactions, admissible operating ranges, power balance relationships, and realistic forms of measurement or control manipulation. The resulting attacks are therefore system-driven rather than detector-driven and can be generated without knowledge of the model. To the best of our knowledge, this broader, physics-aware FDI threat has not been systematically studied for DSGC systems.

\section{System model}
\label{sec:system_model}

\begin{figure}[tb]
    \centering
    \includegraphics[width=.7\columnwidth]{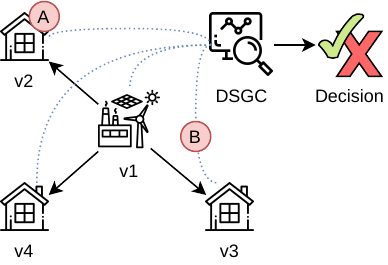}
    \caption{Overview of the system and threat model. An attacker can compromise one of the participants' instrumentation (A) or compromise the communication through substation weaknesses or MitM attacks (B).}
    \label{fig:threat-model}
\end{figure}

We consider a power system based on the DSGC framework~\cite{schafer2015decentral,8587498}. DSGC combines the physical behavior of the grid with a price-based demand-response mechanism. Grid frequency serves as a local indicator of the balance between generation and demand. A frequency drop indicates insufficient generation, whereas a frequency increase reflects excess supply. The electricity price is adjusted according to this signal so that producers and consumers can modify their behavior without relying on centralized price computation~\cite{schafer2015decentral}.
% Let $\theta_i(t)$ denote the phase-angle deviation of participant $i$ from the nominal grid reference, and let $\dot{\theta}_i(t)$ denote its frequency deviation. The dynamics of participant $i$ are given by~\cite{8587498}
% %
% \begin{equation}
% \begin{split}
% \ddot{\theta}_i(t)
% ={}&
% P_i
% -\alpha_i\dot{\theta}_i(t)
% +\sum_{j=1}^{N}
% K_{ij}
% \sin\!\left(
% \theta_j(t)-\theta_i(t)
% \right) \\
% &-
% \frac{\gamma_i}{T_i}
% \left[
% \theta_i(t-\tau_i)
% -
% \theta_i(t-\tau_i-T_i)
% \right],
% \end{split}
% \label{eq:dsgc_dynamics}
% \end{equation}
% %
% where $P_i$ is the nominal power produced or consumed by participant $i$, $\alpha_i$ is the damping coefficient, and $K_{ij}$ represents the coupling between participants $i$ and $j$. The parameter $\gamma_i$ determines the strength of the frequency-dependent price response, $\tau_i$ is the participant's reaction time, and $T_i$ is the interval over which the frequency signal is averaged.

% The final term in~\eqref{eq:dsgc_dynamics} links the economic response of the participants to the physical dynamics of the grid. 
Stability, therefore, depends on the combined effect of the participant reaction times, nominal power values, and price-response coefficients. For each parameter configuration, local stability is assessed around the synchronous operating point through the characteristic equation obtained from the linearized DSGC model~\cite{8587498}.

The system used in this study represents a four participant DSGC system. It consists of one producer, denoted by $v_1$, and three consumers, denoted by $v_2$, $v_3$, and $v_4$. The producer is connected to each consumer, resulting in the star shaped interaction structure, as shown in Figure~\ref{fig:threat-model}. 
% %
% \begin{equation}
% \mathcal{G}
% =
% (\mathcal{V},\mathcal{E}),
% \qquad
% \mathcal{V}
% =
% \{v_1,v_2,v_3,v_4\},
% \end{equation}
%
No direct interaction between the consumer nodes is included in this system representation.
% \begin{figure}
% \centering
% \includegraphics[width=0.75\columnwidth]{Figures/dsgc_star_topology.png}
% \caption{system model} \label{fig7}
% \end{figure} 
Each DSGC configuration is represented as $\mathbf{x}=[\boldsymbol{\tau}^{\top},\mathbf{p}^{\top}, \mathbf{g}^{\top}]^{\top}\in\mathbb{R}^{12}$,
% %
% \begin{equation}
% \mathbf{x}
% =
% [
% \tau_1,\tau_2,\tau_3,\tau_4,
% p_1,p_2,p_3,p_4,
% g_1,g_2,g_3,g_4
% ]^{\top}
% \in\mathbb{R}^{12},
% \label{eq:system_configuration}
% \end{equation}
% %
where $\boldsymbol{\tau},\mathbf{p},\mathbf{g}\in\mathbb{R}^{4}$ denote the reaction-time, power, and price-response variables, respectively. The power values satisfy
\begin{equation}
p_1
=
\left|
p_2+p_3+p_4
\right|,
\label{eq:system_power_balance}
\end{equation}
%
% For the genuine stability experiment, the simulator provides a continuous physical reference value obtained from the roots of the characteristic equation. 
% For a configuration $\mathbf{x}$, this value is defined as
% %
% \begin{equation}
% \sigma(\mathbf{x})
% =
% \max_{\lambda \in \Lambda(\mathbf{x})}
% \operatorname{Re}(\lambda),
% \label{eq:stability_indicator}
% \end{equation}
% %
% where $\Lambda(\mathbf{x})$ denotes the set of characteristic roots associated with the configuration. The sign of $\sigma(\mathbf{x})$ is used only to establish the ground-truth stability label. 
Importantly, characteristic roots are not part of the dataset and are not supplied to the proposed model during training or inference. The stability prediction model receives only the operational parameters $\{\tau_i,p_i,g_i\}_{i=1}^{4}$. 
% and learns a data-driven risk score from stable training configurations and physics-constrained pseudo-negatives.
% StarGNN receives only the operational parameters $\{\tau_i,p_i,g_i\}_{i=1}^{4}$ and learns a data-driven risk score from stable training configurations and physics-constrained pseudo-negatives. 
The detector is designed to identify either genuine physical instability or manipulated system information through the same learned risk score.

\section{Threat Model}
\label{sec:threat_model}
We consider FDI attacks against a deployed DSGC stability monitoring pipeline, as shown in Figure~\ref{fig:threat-model}. The monitored system receives a reported parameter vector $\widetilde{\mathbf{x}}$, defined in Section~\ref{sec:system_model}, and uses it to decide whether the current DSGC configuration should be treated as normal or abnormal. The security property of interest is the integrity of this reported input. If the input is manipulated before it reaches the detector, the stability decision may no longer reflect the actual operating condition.

\paragraph{Attackers capabilities}
The attacker is assumed to compromise part of the measurement or reporting path. Such a compromise may occur through the firmware of microprocessor-based remote terminal units~\cite{10.1145/2994487.2994491}, weaknesses in IEC~61850-based substation communication~\cite{7741376}, or interception of data exchanged between field devices and the control system through a man-in-the-middle attack~\cite{9640002}. The attacker does not need to physically compromise the grid. Instead, the attacker tampers with the values that describe the DSGC configuration before those values are processed by the monitoring system.

We consider an integrity attacker with partial control over the reported parameters. Depending on the scenario, the adversary may alter a single parameter, several parameters belonging to one participant, or parameters reported by multiple participants. The attacker may modify reaction-time parameters, reported power values, price-response coefficients, or combinations of these features. This captures both localized compromises, such as a single reporting unit, and coordinated compromises affecting several reported attributes.
The attacks are performed at inference time. The adversary is not assumed to modify the training data, the learned detector parameters, the classification threshold, or the stability labels.
The focus is instead on data integrity attacks that provide a syntactically valid but manipulated parameter vector.

We consider four attacker capability profiles, as summarized in Table~\ref{tab:attacker_profiles}. Many of the Generic non adaptive FDI attacks are black-box with respect to the detector, as they require no access to its architecture, parameters, gradients, decision threshold, or anomaly score. Balance-preserving FDI attacks follow a grey-box setting in which the attacker has limited knowledge of the DSGC physical model, including the power-balance relation $p_1=\left|p_2+p_3+p_4\right|$ but has no internal knowledge of the detector. Replay-style FDI attacks represent a second grey-box capability in which the attacker can observe and reuse previously valid participant reports.
We also consider a constrained adaptive grey-box attacker (single node adaptive concealment attacker) that targets instability concealment. The attacker compromises a single participant and can modify only its reported reaction time and price response parameters, $\tau_i$ and $g_i$, while all power measurements remain unchanged. It has no access to the model architecture, parameters, gradients, continuous anomaly score, or decision threshold, but it can submit a limited number of candidates and observe the corresponding binary alarm response.

\begin{table}[t]
\centering
\footnotesize
\caption{Attacker capability profiles.}
\label{tab:attacker_profiles}
\begin{tabular}{lll}
\toprule
\textbf{Attacker} & \textbf{Access} & \textbf{Capability} \\
\midrule
Generic non-adaptive & Black-box & Modify $\tau,p,g$ reports \\
Balance-preserving & Grey-box & Knows power-balance relation \\
Replay style & Grey-box & Reuses valid reports \\
Adaptive single node & Grey-box & Modifies $\tau_i,g_i$; binary feedback \\
\bottomrule
\end{tabular}
\end{table}

\paragraph{Attacker Objective.} The main attacker objective is the \textit{concealment} of an unreliable or manipulated configuration to be accepted as normal. In the most safety-critical case, an actually unstable configuration is reported in a way that makes it appear stable, which may delay or suppress corrective actions~\cite{efatinasab2024gangrid,EFATINASAB2025101799}. In severe cases, this may allow disturbances to propagate and contribute to cascading failures or wider outages, with consequences for dependent services such as telecommunications, transportation, and emergency response~\cite{wickerson2024gridfailure}. 

The second objective is to create \textit{false alarms} by making stable configurations appear abnormal. Such false alarms can reduce operator confidence in the monitoring system and may trigger unnecessary interventions.
The detector uses a single abnormal class. Both genuinely unstable DSGC configurations and attacked parameter reports are mapped to this class, since manipulated reports cannot be trusted as evidence of stable operation. This conservative design provides an operational warning that corrective action is needed since either case invalidates the assumption that the reported configuration represents safe operation, particularly when an attacker attempts to conceal an unstable condition. It also remains effective for attacks that create false instability alarms, as both cases indicate that the reported grid state is unreliable and needs further investigation.

None of the attack samples is used to train the detector or to construct the physics-constrained pseudo-negative samples. The evaluation, therefore, tests whether the detector can identify previously unseen manipulations using only stable training data and the physical structure encoded in the proposed method.

\section{False Data Injection Attacks}
\label{sec:fdi}

Based on the capabilities of different attackers, we defined and evaluated nine non-adaptive FDI attacks that cover parameter bias, power manipulation, participant compromise, coordinated manipulation, load redistribution, and replay style substitution. Table~\ref{tab:attack_summary} summarizes the attacker scope and the system constraints preserved by each scenario. The attacks are generated only at inference time and are independent of the StarGNN detector. All manipulated values are restricted to admissible feature ranges derived from the clean training data. Detailed perturbation magnitudes and generation procedures are reported in Appendix~\ref{app:attack_details}.

The first six attacks provide generic DSGC-specific manipulations over reaction time, price response, and power reports. The remaining three adapt established attack patterns from the power system security literature: load redistribution, localized load redistribution, and replay. These scenarios range from isolated parameter falsification to coordinated and constraint-preserving manipulation. The attack IDs in Table~\ref{tab:attack_summary} map to the capability profiles in Table~\ref{tab:attacker_profiles} as follows: A1--A3 and A5 use the generic non-adaptive black-box setting; A4, A7, and A8 use the balance-preserving grey-box setting; A9 follows the replay style grey-box setting; and A10 uses he adaptive single-node grey-box setting. A6 can follow either the generic black-box or balance-preserving grey-box profile depending on whether its power component is based on A3 or A4.
\begin{table*}[t]
\centering
\footnotesize
\caption{Summary of the evaluated FDI attacks. Detailed perturbation rules and
parameter ranges are provided in Appendix~\ref{app:attack_details}.}
\label{tab:attack_summary}
\renewcommand{\arraystretch}{1.10}

\begin{tabular}{c p{2.8cm}p{2.5cm}p{2.5cm}p{3.7cm}p{3.7cm}}
\toprule
\textbf{ID}
& \textbf{Attack}
& \textbf{Modified reports}
& \textbf{Compromise scope}
& \textbf{Constraint / characteristic}
& \textbf{Motivation} \\
\midrule

A1
& Reaction-time bias
& $\tau$
& 1--4 parameters
& Bounded additive bias
& Time-delay-inspired manipulation~\cite{10.1145/3055386.3055392} \\

A2
& Price-response bias
& $g$
& 1--4 parameters
& Bounded additive bias
& Falsification of reported price-response behavior \\

A3
& Power imbalance
& Consumer $p$
& 1--3 consumers
& Producer power remains unchanged
& Load-altering attack~\cite{5976424} \\

A4
& Balance-preserving power
& Consumer and producer $p$
& 1--3 consumers
& Producer power recomputed to preserve power balance
& Stealthy FDI through constraint-preserving manipulation \\

A5
& Node takeover
& $\tau,p,g$
& One participant
& Complete local report is modified
& Compromise of one participant's reporting unit \\

A6
& Coordinated FDI
& $\tau,g,p$
& Multiple feature groups
& Reaction-time, price-response, and power reports are jointly modified
& Coordinated falsification across feature groups \\

A7
& Load redistribution
& Consumer $p$
& Three consumers
& Aggregate consumer demand preserved before clipping; producer power recomputed
& Load-redistribution attack~\cite{10521762} \\

A8
& Local load redistribution
& Consumer $p$
& 1--2 consumers
& Total consumer demand approximately preserved; producer power recomputed
& Local load-redistribution attack~\cite{6805238} \\

A9
& Replay style node substitution
& $\tau,p,g$
& One participant
& Participant report replaced using values from another stable configuration
& Static adaptation of replay attacks~\cite{5394956} \\

A10
& Single node adaptive concealment
& $\tau,g$
& One participant
& Power unchanged; $50\%$ range bound; 25 additional binary-alarm queries
& Adaptive instability concealment \\

\bottomrule
\end{tabular}
\end{table*}

We additionally consider a constrained adaptive attacker whose objective is to conceal an unstable operating point. The adversary compromises one participant and modifies only its reaction time and price response reports, $\tau_i$ and $g_i$, while all power reports remain unchanged. Each mutable feature may deviate by at most $50\%$ of its training-derived range and remains within its admissible bounds. The attacker has no access to the StarGNN architecture, parameters, continuous risk score, or decision threshold. It can only submit candidate reports and observe the corresponding binary alarm. Starting from an initial random perturbation, it performs a bounded randomized search using 25 additional queries, for a maximum of 26 detector evaluations per sample. The search combines local exploration around the current candidate with global sampling within the admissible perturbation region. Among alarm free candidates, the attacker retains the one with the largest normalized displacement from the original configuration. Further implementation details are provided in Appendix~\ref{app:attack_details}.

\section{Methodology}\label{sec:methodology}

%\subsection{Framework Overview}
%\label{subsec:framework_overview}

\begin{figure}
    \centering
    \includegraphics[width=\linewidth]{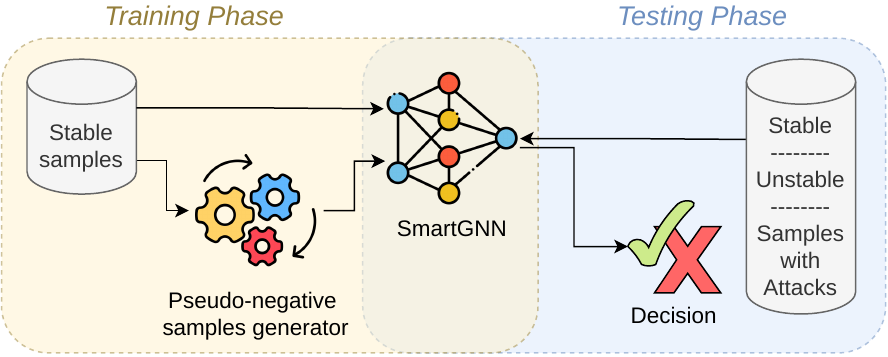}
    \caption{Overview of training and testing of StarGNN.}
    \label{fig:pipeline}
\end{figure}

The proposed training and testing pipeline of StarGNN is represented in Figure~\ref{fig:pipeline}. Stable configuration samples are used to generate physics-constrained pseudo-negatives, after which both are represented as producer--consumer star graphs and processed by StarGNN. The model learns a scalar risk score through separation, ranking, compactness, and variance-based objectives. A threshold calibrated solely from stable validation scores is then used for both instability prediction and detection of unseen FDI attacks. A representation of the different layers of the proposed step is shown in the Appendix, Figure~\ref{fig:framework}.

\subsection{Stable-Only Data Preparation}
\label{subsec:stable_data_preparation}

The dataset is first divided according to the binary stability label. Let
$\mathcal{D}_{s}$ and $\mathcal{D}_{u}$ denote the sets of stable and unstable configurations, respectively. Only samples in $\mathcal{D}_{s}$ are used during model development. The stable configurations are randomly divided into three non-overlapping partitions, where $60\%$ of the stable samples are assigned to training, $20\%$ to validation, and the remaining $20\%$ to testing. All unstable configurations are excluded from the training and validation stages. Instead, the complete unstable subset is retained for final evaluation.
The final stability prediction test set is formed by combining the held-out stable samples with all available unstable samples. The stable test partition is also kept separately because unseen attacks are generated only from held-out stable configurations.

Feature standardization is carried out using statistics computed only from the stable training partition. 
% For feature $k$, the standardized value is
% %
% \begin{equation}
% \bar{x}_{k}
% =
% \frac{x_k-\mu_k^{\mathrm{tr}}}
% {\max\!\left(s_k^{\mathrm{tr}},\epsilon\right)},
% \label{eq:stable_standardization}
% \end{equation}
% %
% where $\mu_k^{\mathrm{tr}}$ and $s_k^{\mathrm{tr}}$ are the mean and standard deviation of feature $k$ over $\mathcal{D}_{s}^{\mathrm{tr}}$, and $\epsilon=10^{-8}$ prevents numerical instability.
The same statistics are then applied to the stable validation set, the final test set, and all attacked samples. No information from unstable or attacked configurations is used to fit the scaler.
The minimum and maximum values of each feature are also estimated from the stable training partition. These empirical ranges are used only to define bounded perturbation regions for the later generation of pseudo-negatives and attack samples. For feature $k$, the expanded interval is written as
\begin{equation}
\left[
x_{k,\min}^{\mathrm{tr}}
-
\delta R_k,
\;
x_{k,\max}^{\mathrm{tr}}
+
\delta R_k
\right],
\label{eq:expanded_bounds}
\end{equation}
where $\delta=0.25$ and 
\begin{equation}
R_k
=
\max\!\left(
x_{k,\max}^{\mathrm{tr}}
-
x_{k,\min}^{\mathrm{tr}},
10^{-6}
\right)
\end{equation}

\subsection{Physics-Constrained Pseudo-Negative Generation}
\label{subsec:pseudo_negative_generation}

Training only on stable configurations provides no direct indication of how the risk score should behave away from the observed stable data. To introduce such contrast without using genuine unstable samples, we generate a set of pseudo-negative configurations from each stable training sample. These samples are used to shape the learned decision boundary, but they are not treated as physically verified unstable operating points.
Let
\begin{equation}
\mathbf{x}
=
[
\tau_1,\ldots,\tau_4,
p_1,\ldots,p_4,
g_1,\ldots,g_4
]^{\top}
\in
\mathcal{D}_{s}^{\mathrm{tr}}
\end{equation}
denote a stable training configuration. A pseudo-negative
$\widetilde{\mathbf{x}}$ is obtained by perturbing selected reaction time parameters, price response coefficients, or both. The nominal power values are not intentionally perturbed during this stage.
Three perturbation types are considered. A price response perturbation is selected with probability $2/5$, a reaction time perturbation with probability $1/5$, and a combined perturbation with probability $2/5$.
For each selected parameter group, between one and four participant variables are chosen uniformly without replacement.

For a selected price response coefficient $g_i$, a direction
$d_i\in\{-1,+1\}$ is sampled with equal probability. The modified value is
\begin{equation}
\begin{aligned}
\widetilde{g}_i
&=
g_i+d_i\alpha_i D_i^{g},
&
\alpha_i
&\sim
\mathcal{U}(0.20,0.65).
\end{aligned}
\label{eq:g_pseudo_perturbation}
\end{equation}
% %
% \begin{equation}
% \widetilde{g}_i
% =
% g_i
% +
% d_i\alpha_i D_i^{g},
% \label{eq:g_pseudo_perturbation}
% \end{equation}
% %
% where
% %
% \begin{equation}
% \alpha_i
% \sim
% \mathcal{U}(0.20,0.65)
% \end{equation}
% %
and $D_i^{g}$ is the available distance from the current value to the corresponding parameter limit:
\begin{equation}
D_i^{g}
=
\begin{cases}
g_{\max}-g_i, & d_i=+1,\\
g_i-g_{\min}, & d_i=-1.
\end{cases}
\end{equation}
The admissible interval is $g_i \in [0.05,1.0]$.
%
% \begin{equation}
% g_i\in[0.05,1.0].
% \end{equation}
Reaction time perturbations are generated in the same way. For a selected
$\tau_i$,
\begin{equation}
\begin{aligned}
\widetilde{\tau}_i
&=
\tau_i+d_i\beta_i D_i^{\tau},
&
\beta_i
&\sim
\mathcal{U}(0.10,0.35).
\end{aligned}
\label{eq:tau_pseudo_perturbation}
\end{equation}

% %
% \begin{equation}
% \widetilde{\tau}_i
% =
% \tau_i
% +
% d_i\beta_i D_i^{\tau},
% \label{eq:tau_pseudo_perturbation}
% \end{equation}
% %
% where
% %
% \begin{equation}
% \beta_i
% \sim
% \mathcal{U}(0.10,0.35)
% \end{equation}
% %
and
\begin{equation}
D_i^{\tau}
=
\begin{cases}
\tau_{\max}-\tau_i, & d_i=+1,\\
\tau_i-\tau_{\min}, & d_i=-1.
\end{cases}
\end{equation}
The reaction times are restricted to $\tau_i \in [0.5,10.0]$.
%
% \begin{equation}
% \tau_i\in[0.5,10.0].
% \end{equation}
%

For a combined perturbation, the price response and reaction time
transformations are applied sequentially.
The perturbations are non-directional. In particular, the generator does not assume that increasing a reaction time or price response coefficient always moves the system toward instability. Both upward and downward changes are considered so that the model is not tied to a predetermined monotonic relation between a parameter and the stability condition. After each transformation, the variables are clipped to their admissible
domains: $\tau_i\in[0.5,10.0]$, $g_i\in[0.05,1.0]$,
$p_i\in[-2.0,-0.5]$ for $i\in\{2,3,4\}$, and
$p_1\in[1.5,6.0]$.

% After each transformation, the variables are clipped to their admissible
% domains:
% %
% \begin{equation}
% \tau_i\in[0.5,10.0],
% \qquad
% g_i\in[0.05,1.0],
% \end{equation}
% %
% \begin{equation}
% p_i\in[-2.0,-0.5],
% \qquad i\in\{2,3,4\},
% \end{equation}
% %
% and
% %
% \begin{equation}
% p_1\in[1.5,6.0].
% \end{equation}
% %
The producer power is then projected onto the producer--consumer balance
constraint $\widetilde{p}_1 =
\left|\widetilde{p}_2+\widetilde{p}_3+\widetilde{p}_4\right|$.
%
% \begin{equation}
% \widetilde{p}_1
% =
% \left|
% \widetilde{p}_2+
% \widetilde{p}_3+
% \widetilde{p}_4
% \right|.
% \label{eq:pseudo_power_projection}
% \end{equation}
%
This step ensures that the pseudo-negatives do not rely on an obvious power-balance violation and remain consistent with the basic structure of the DSGC data.
For every stable training configuration, $K=16$ independently generated variants are stored: $\mathcal{P}(\mathbf{x}_n)=\{\widetilde{\mathbf{x}}_{n}^{(1)},\ldots,\widetilde{\mathbf{x}}_{n}^{(K)}\}$.
%
% \begin{equation}
% \mathcal{P}(\mathbf{x}_n)
% =
% \left\{
% \widetilde{\mathbf{x}}_{n}^{(1)},
% \ldots,
% \widetilde{\mathbf{x}}_{n}^{(K)}
% \right\}.
% \end{equation}
%
% The complete pseudo-negative bank therefore has the form
% %
% \begin{equation}
% \mathcal{P}
% \in
% \mathbb{R}^{N_s\times K\times12},
% \end{equation}
%
% where $N_s$ is the number of stable training samples. 
During each training iteration, one of the $K$ variants is selected at random for every stable
sample in the batch. This allows the same stable configuration to be paired with different perturbations over the course of training without increasing the batch size.
The generated samples receive a pseudo-negative target only for the purpose of contrastive boundary learning.
They should therefore not be interpreted as synthetic ground-truth unstable samples. Their role is to expose the model to admissible departures from observed stable configurations while preserving the main parameter limits and the producer--consumer power relation.

\subsection{Topology-Aware Graph Representation}
\label{subsec:graph_representation}

The original dataset stores each DSGC configuration as a twelve dimensional tabular vector. Although this form is convenient for data storage, it does not explicitly preserve the producer--consumer structure of the system. We therefore reorganize every configuration as a graph whose nodes correspond to the four DSGC participants.
% Let
% %
% \begin{equation}
% \mathbf{x}
% =
% [
% \tau_1,\ldots,\tau_4,
% p_1,\ldots,p_4,
% g_1,\ldots,g_4
% ]^{\top}
% \in\mathbb{R}^{12}
% \end{equation}
% %
% denote an input configuration. 
Before graph construction, each feature is standardized using the mean and standard deviation obtained from the stable training partition, as described in Section~\ref{subsec:stable_data_preparation}. 
% The standardized configuration is denoted by
% %
% \begin{equation}
% \overline{\mathbf{x}}
% =
% [
% \overline{\tau}_1,\ldots,\overline{\tau}_4,
% \overline{p}_1,\ldots,\overline{p}_4,
% \overline{g}_1,\ldots,\overline{g}_4
% ]^{\top}.
% \end{equation}
The three parameters belonging to participant $i$ are grouped into a node feature vector:
\begin{equation}
\mathbf{h}_i^{(0)}
=
\left[
\overline{\tau}_i,
\overline{p}_i,
\overline{g}_i
\right]^{\top},
\qquad
i\in\{1,2,3,4\}.
\label{eq:graph_node_features}
\end{equation}
Here, $\overline{\tau}_i$ represents the standardized reaction time, $\overline{p}_i$ the standardized nominal power, and $\overline{g}_i$ the standardized price response coefficient of participant $i$.
The four node vectors are stacked to form the initial node feature matrix
\begin{equation}
\mathbf{H}^{(0)}
=
\begin{bmatrix}
(\mathbf{h}_1^{(0)})^{\top}\\
(\mathbf{h}_2^{(0)})^{\top}\\
(\mathbf{h}_3^{(0)})^{\top}\\
(\mathbf{h}_4^{(0)})^{\top}
\end{bmatrix}
\in\mathbb{R}^{4\times3}.
\label{eq:graph_feature_matrix}
\end{equation}
The first row corresponds to the producer, while the remaining three rows correspond to the consumers.
The node feature matrix is associated with the star-shaped graph introduced in Section~\ref{sec:system_model}.
\begin{equation}
\begin{aligned}
\mathcal{G}
&=
(\mathcal{V},\mathcal{E}),
&
\mathcal{V}
&=
{v_1,v_2,v_3,v_4}.
\end{aligned}
\end{equation}

% %
% \begin{equation}
% \mathcal{G}
% =
% (\mathcal{V},\mathcal{E}),
% \end{equation}
% %
% where
% %
% \begin{equation}
% \mathcal{V}
% =
% \{v_1,v_2,v_3,v_4\}
% \end{equation}
% %
and
\begin{equation}
\mathcal{E}
=
\left\{
(v_1,v_2),
(v_1,v_3),
(v_1,v_4)
\right\}.
\end{equation}
Node $v_1$ represents the producer, and nodes $v_2$, $v_3$, and $v_4$ represent the consumers. The graph, therefore, preserves the producer--consumer interaction pattern contained in the DSGC model rather than treating the twelve input variables as unrelated features.
The same transformation is applied to stable samples, pseudo-negative configurations, genuine unstable samples, and attacked observations.

\subsection{StarGNN Risk Model}
\label{subsec:stargnn}

The proposed model follows the general message passing neural network framework, in which a node updates its representation using its own features and information received from neighboring nodes~\cite{gilmer2017neural}. We adapt this idea to the fixed producer--consumer structure of the DSGC system.
The resulting architecture, referred to as StarGNN, uses role-aware message passing and graph aggregation. The producer receives information from all three consumers, while each consumer receives information only from the producer. The producer representation is also retained separately when the node embeddings are combined at the graph level. Standard operations such as linear transformations, activation functions, normalization, dropout, and pooling are used throughout the network. The DSGC-specific contribution lies in how these operations are arranged to reflect the different roles of the producer and consumers.

The model consists of a shared node encoder, two star structured message-passing layers, a graph level aggregation stage, and two output heads. One head produces a graph embedding used by the stable-only training objective, while the other produces the scalar risk score used for detection.
Let
\begin{equation}
\mathbf{H}^{(0)}
=
\left[
\mathbf{h}_1^{(0)},
\mathbf{h}_2^{(0)},
\mathbf{h}_3^{(0)},
\mathbf{h}_4^{(0)}
\right]^{\top}
\in\mathbb{R}^{4\times3}
\end{equation}
denote the node feature matrix of one DSGC configuration. Each row contains the standardized reaction time, nominal power, and price response coefficient of one participant.
A shared multilayer encoder maps each three-dimensional node vector to the hidden space:
\begin{equation}
\mathbf{h}_i^{(0,e)}
=
\operatorname{ReLU}
\left(
\mathbf{W}_{e,2}
\operatorname{ReLU}
\left(
\mathbf{W}_{e,1}\mathbf{h}_i^{(0)}
+
\mathbf{b}_{e,1}
\right)
+
\mathbf{b}_{e,2}
\right).
\label{eq:node_encoder}
\end{equation}
The same encoder parameters are used for the producer and the three consumers. The hidden dimension is set to $d_h=128$.
The encoded node representations are then processed by $L=2$ message-passing layers. At layer $\ell$, the transformed message of node $i$ is
\begin{equation}
\mathbf{u}_i^{(\ell)}
=
\mathbf{W}_{m}^{(\ell)}
\mathbf{h}_i^{(\ell)}
+
\mathbf{b}_{m}^{(\ell)}.
\label{eq:node_message}
\end{equation}
The producer receives the mean of the three consumer messages:
\begin{equation}
\mathbf{m}_1^{(\ell)}
=
\frac{1}{3}
\sum_{j=2}^{4}
\mathbf{u}_j^{(\ell)}.
\label{eq:producer_message}
\end{equation}
Each consumer receives the message generated by the producer:
\begin{equation}
\mathbf{m}_i^{(\ell)}
=
\mathbf{u}_1^{(\ell)},
\qquad
i\in\{2,3,4\}.
\label{eq:consumer_message}
\end{equation}
The node update combines the incoming message with a separate transformation of the current node representation:
\begin{equation}
\mathbf{h}_i^{(\ell+1)}
=
\operatorname{Dropout}
\left(
\operatorname{LN}
\left[
\operatorname{ReLU}
\left(
\mathbf{W}_{s}^{(\ell)}
\mathbf{h}_i^{(\ell)}
+
\mathbf{b}_{s}^{(\ell)}
+
\mathbf{m}_i^{(\ell)}
\right)
\right]
\right),
\label{eq:stargnn_update}
\end{equation}
where $\mathbf{W}_{s}^{(\ell)}$ and $\mathbf{W}_{m}^{(\ell)}$ are trainable self and message transformations, respectively. The operator $\operatorname{LN}(\cdot)$ denotes layer normalization. A dropout rate of $0.10$ is applied after each message-passing layer.

This communication rule follows the DSGC star structure directly. The producer gathers information from all consumers, while the consumers do not exchange messages with one another. So we avoid introducing edges that are not present in the adopted system model.
After the final message-passing layer, the producer embedding is retained as
\begin{equation}
\mathbf{h}_{p}
=
\mathbf{h}_1^{(L)}.
\end{equation}
The three consumer embeddings are summarized using mean and element-wise
maximum pooling:
\begin{equation}
\begin{aligned}
\mathbf{h}_{c}^{\mathrm{mean}}
&=
\frac{1}{3}\sum_{i=2}^{4}\mathbf{h}_i^{(L)},
&
\mathbf{h}_{c}^{\mathrm{max}}
&=
\max_{i\in\{2,3,4\}}\mathbf{h}_i^{(L)}.
\end{aligned}
\label{eq:consumer_pooling}
\end{equation}
%
% The three consumer embeddings are summarized using mean pooling,
% %
% \begin{equation}
% \mathbf{h}_{c}^{\mathrm{mean}}
% =
% \frac{1}{3}
% \sum_{i=2}^{4}
% \mathbf{h}_i^{(L)},
% \label{eq:consumer_mean_pool}
% \end{equation}
% %
% and element-wise maximum pooling,
% %
% \begin{equation}
% \mathbf{h}_{c}^{\mathrm{max}}
% =
% \max_{i\in\{2,3,4\}}
% \mathbf{h}_i^{(L)}.
% \label{eq:consumer_max_pool}
% \end{equation}
% %
The graph level representation is then formed as
\begin{equation}
\mathbf{q}
=
\mathbf{h}_{p}
\mathbin{\Vert}
\mathbf{h}_{c}^{\mathrm{mean}}
\mathbin{\Vert}
\mathbf{h}_{c}^{\mathrm{max}}
\in\mathbb{R}^{3d_h},
\label{eq:graph_representation}
\end{equation}
where $\Vert$ denotes concatenation.
This role-aware aggregation differs from a conventional global mean or maximum pooling operation, which treats all nodes in the same way. Keeping the producer embedding separate preserves its central role in the DSGC system. The mean consumer embedding represents their shared behavior, while maximum pooling retains the strongest response observed among the three consumers.

A graph head maps $\mathbf{q}$ to a lower-dimensional embedding:
\begin{equation}
\mathbf{z}_{\phi}(\mathbf{x})
=
\mathbf{W}_{z,2}
\operatorname{Dropout}
\left[
\operatorname{ReLU}
\left(
\mathbf{W}_{z,1}\mathbf{q}
+
\mathbf{b}_{z,1}
\right)
\right]
+
\mathbf{b}_{z,2},
\label{eq:graph_embedding}
\end{equation}
where  $\mathbf{z}_{\phi}(\mathbf{x}) \in \mathbb{R}^{d_z}$.
% %
% \begin{equation}
% \mathbf{z}_{\phi}(\mathbf{x})
% \in\mathbb{R}^{d_z}
% \end{equation}
% %
and $d_z=64$. This embedding is used by the compactness and variance terms introduced in Section~\ref{subsec:training_objective}.
The risk maps the graph embedding to a scalar logit:
\begin{equation}
r_{\phi}(\mathbf{x})
=
\mathbf{w}_{r,2}^{\top}
\operatorname{Dropout}
\left[
\operatorname{ReLU}
\left(
\mathbf{W}_{r,1}
\mathbf{z}_{\phi}(\mathbf{x})
+
\mathbf{b}_{r,1}
\right)
\right]
+
b_{r,2}.
\label{eq:stargnn_risk}
\end{equation}
The logit is used directly as the risk score and is not converted into a probability at inference time. A larger value indicates that the input provides a stronger instability cue relative to the stable configurations observed during training.
The model, therefore, returns
\begin{equation}
f_{\phi}(\mathbf{x},\mathcal{G})
=
\left(
\mathbf{z}_{\phi}(\mathbf{x}),
r_{\phi}(\mathbf{x})
\right).
\label{eq:stargnn_outputs}
\end{equation}

\subsection{Stable-Only Training Objective}
\label{subsec:training_objective}

The StarGNN is trained using stable configurations and the pseudo-negatives derived from them. For each stable sample in a mini-batch, one of its precomputed pseudo-negative variants is selected at random. The clean configuration and the selected variant are converted to graphs and passed through the same network.
For a mini-batch of size $B$, let
\begin{equation}
\begin{aligned}
\left(\mathbf{z}_{n}^{s}, r_{n}^{s}\right)
&= f_{\phi}\left(\mathbf{x}_{n},\mathcal{G}\right),
&
\left(\mathbf{z}_{n}^{p}, r_{n}^{p}\right)
&= f_{\phi}\left(\widetilde{\mathbf{x}}_{n},\mathcal{G}\right).
\end{aligned}
\end{equation}
The first pair denotes the embedding and risk score of the stable sample
$\mathbf{x}_{n}$, whereas the second corresponds to the selected
pseudo-negative $\widetilde{\mathbf{x}}_{n}$.

% %
% \begin{equation}
% \left(
% \mathbf{z}_{n}^{s},
% r_{n}^{s}
% \right)
% =
% f_{\phi}
% \left(
% \mathbf{x}_{n},
% \mathcal{G}
% \right)
% \end{equation}
% %
% denote the embedding and risk score of stable sample $\mathbf{x}_{n}$, and let
% %
% \begin{equation}
% \left(
% \mathbf{z}_{n}^{p},
% r_{n}^{p}
% \right)
% =
% f_{\phi}
% \left(
% \widetilde{\mathbf{x}}_{n},
% \mathcal{G}
% \right)
% \end{equation}
% %
% denote the corresponding outputs for the selected pseudo-negative
% $\widetilde{\mathbf{x}}_{n}$. 
The training objective contains four terms:
a binary separation loss, a pairwise ranking loss, a stable-embedding
compactness term, and a variance regularizer.
The binary separation term assigns target $0$ to stable samples and target
$1$ to pseudo-negatives:
\begin{equation}
\begin{split}
\mathcal{L}_{\mathrm{BCE}}
=
-\frac{1}{2B}
\sum_{n=1}^{B}
\Big[
&
\log
\left(
1-\operatorname{sigmoid}(r_n^{s})
\right)
\\
&+
\log
\left(
\operatorname{sigmoid}(r_n^{p})
\right)
\Big].
\end{split}
\label{eq:bce_loss}
\end{equation}
The pseudo-negative target is used only to provide contrast during
training.
The binary term separates the two groups at the batch level, but it does
not explicitly preserve the pairing between a stable configuration and
the variant generated from it. We therefore add a ranking loss:
\begin{equation}
\mathcal{L}_{\mathrm{rank}}
=
\frac{1}{B}
\sum_{n=1}^{B}
\max
\left(
0,
m+r_n^{s}-r_n^{p}
\right),
\label{eq:ranking_loss}
\end{equation}
where $m=1$ is the ranking margin. This term encourages the pseudo-negative
associated with each stable configuration to receive a risk score at
least $m$ units higher than the score of its clean counterpart.
In addition to score separation, the stable embeddings are encouraged to
form a compact region. Following the one class representation learning principle used in Deep SVDD~\cite{ruff2018deep}, a center
$\mathbf{c}\in\mathbb{R}^{d_z}$ is calculated from the embeddings of all stable training samples:
\begin{equation}
\mathbf{c}
=
\frac{1}{N_s}
\sum_{n=1}^{N_s}
\mathbf{z}_{\phi}
\left(
\mathbf{x}_n
\right),
\qquad
\mathbf{x}_n
\in
\mathcal{D}_{s}^{\mathrm{tr}}.
\label{eq:svdd_center}
\end{equation}
The center is treated as a fixed, non-trainable quantity while the network
parameters are updated. The compactness loss is
\begin{equation}
\mathcal{L}_{\mathrm{SVDD}}
=
\frac{1}{B}
\sum_{n=1}^{B}
\left\|
\mathbf{z}_{n}^{s}
-
\mathbf{c}
\right\|_{2}^{2}.
\label{eq:svdd_loss}
\end{equation}
Only stable embeddings contribute to this term. Pseudo-negative embeddings are not forced toward the stable center.
A compactness objective alone can reduce the diversity of the learned representation. To avoid all stable embeddings collapsing to nearly the same point, we impose a minimum standard deviation on every embedding dimension. This follows the variance-floor idea used to prevent representation collapse in variance-regularized learning~\cite{bardes2022vicreg}. Let
\begin{equation}
s_k
=
\sqrt{
\operatorname{Var}
\left(
z_{1k}^{s},
\ldots,
z_{Bk}^{s}
\right)
+
\epsilon
}
\end{equation}
denote the batch standard deviation of embedding dimension $k$. The  variance penalty is
\begin{equation}
\mathcal{L}_{\mathrm{var}}
=
\frac{1}{d_z}
\sum_{k=1}^{d_z}
\max
\left(
0,
s_0-s_k
\right),
\label{eq:variance_loss}
\end{equation}
where $s_0=0.5$ and $\epsilon=10^{-6}$. Dimensions whose standard
deviation is already above the target value, receive no penalty.
The complete objective is
\begin{equation}
\mathcal{L}
=
\lambda_{\mathrm{BCE}}
\mathcal{L}_{\mathrm{BCE}}
+
\lambda_{\mathrm{rank}}
\mathcal{L}_{\mathrm{rank}}
+
\lambda_{\mathrm{SVDD}}
\mathcal{L}_{\mathrm{SVDD}}
+
\lambda_{\mathrm{var}}
\mathcal{L}_{\mathrm{var}},
\label{eq:total_training_loss}
\end{equation}
The loss weights were set to $\lambda_{\mathrm{BCE}}=1.0$, $\lambda_{\mathrm{rank}}=0.5$, $\lambda_{\mathrm{SVDD}}=0.05$, and $\lambda_{\mathrm{var}}=0.10$.
The stable center is initialized from the embeddings produced by the initial network and is recomputed every ten training epochs. This periodic update allows the center to follow the developing stable representation without treating it as a trainable model parameter.
The model is trained for $150$ epochs using AdamW with a learning rate of $10^{-3}$, weight decay of $10^{-5}$, and a mini-batch size of $128$. Gradient norms are clipped at $5$ during optimization. The detector uses only the raw StarGNN risk logit $r_{\phi}(\mathbf{x})$, while the SVDD and
variance terms act as regularizer during training.
% \subsection{Threshold Calibration and Inference}
% \label{subsec:threshold_inference}
After training, the model produces a scalar risk logit
$r_{\phi}(\mathbf{x})$ for each input configuration. This raw logit is
used directly as the detection score. No sigmoid transformation or SVDD
distance is added during inference.
The decision threshold is calibrated using only the held-out stable validation set:
\begin{equation}
\begin{aligned}
\mathcal{R}_{s}^{\mathrm{val}}
&=
\left\{
r_{\phi}(\mathbf{x}) :
\mathbf{x}\in\mathcal{D}_{s}^{\mathrm{val}}
\right\},
&
\eta
&=
Q_q\!\left(\mathcal{R}_{s}^{\mathrm{val}}\right).
\end{aligned}
\label{eq:risk_threshold}
\end{equation}
Here, $Q_q(\cdot)$ denotes the empirical quantile operator, and $q=0.90$ in the implemented configuration.

Because the threshold is selected from stable validation scores alone, neither genuine unstable configurations nor attacked samples influence
its value. The final test set is used only after the model and threshold have been fixed.

The same risk score and decision threshold are used for both evaluation tasks. In the stability prediction experiment, samples above the threshold are classified as unstable. In the attack detection part, a manipulated report above the same threshold is flagged as containing an instability cue. 
% No attack specific score, threshold, or recalibration is
% introduced. 
The use of the $0.90$ quantile corresponds to a relatively permissive operating point. In the absence of tied scores, approximately $10\%$ of the stable validation configurations lie above the threshold.

\section{Evaluation}
\label{sec:evaluation}

% \subsection{Evaluation Metrics}
% \label{subsec:evaluation_metrics}

In this section, we evaluate StarGNN. First, we present the dataset used in Section~\ref{subsec:dataset}, which contains both stable and unstable data but no attacks. Then, Section~\ref{subsec:attack_impact} investigates the impact of the attacks introduced in Section~\ref{sec:fdi} on the dataset. Section~\ref{subsec:stability_results} showcases the capabilities of StarGNN in detecting unstable configurations, while Section~\ref{subsec:attack_results} discusses the capabilities in identifying FDI attacks.

The reported metrics for evaluation of our approach are accuracy, macro-averaged precision, macro-averaged recall, macro-averaged F1-score, and the area under the receiver operating characteristic curve. Let $\mathrm{TP}$, $\mathrm{TN}$, $\mathrm{FP}$, and $\mathrm{FN}$ denote the numbers of true positives, true negatives, false positives, and false negatives, respectively. Accuracy measures the proportion of correctly classified samples:
\begin{equation}
\mathrm{Accuracy}
=
\frac{\mathrm{TP}+\mathrm{TN}}
{\mathrm{TP}+\mathrm{TN}+\mathrm{FP}+\mathrm{FN}}.
\label{eq:accuracy}
\end{equation}

For each class $c\in\{0,1\}$, precision and recall are calculated as
\begin{equation}
\mathrm{Precision}_{c}
=
\frac{\mathrm{TP}_{c}}
{\mathrm{TP}_{c}+\mathrm{FP}_{c}},
\qquad
\mathrm{Recall}_{c}
=
\frac{\mathrm{TP}_{c}}
{\mathrm{TP}_{c}+\mathrm{FN}_{c}}.
\label{eq:class_precision_recall}
\end{equation}

The class-specific F1-score is the harmonic mean of precision and recall:
\begin{equation}
\mathrm{F1}_{c}
=
2
\frac{
\mathrm{Precision}_{c}\mathrm{Recall}_{c}
}{
\mathrm{Precision}_{c}+\mathrm{Recall}_{c}
}.
\label{eq:class_f1}
\end{equation}
Because the two classes may contain different numbers of samples, precision, recall, and F1-score are reported using macro averaging, so that each class contributes equally. For a metric
$M\in\{\mathrm{Precision},\mathrm{Recall},\mathrm{F1}\}$,
the macro-averaged value is computed as
\begin{equation}
\mathrm{Macro}\text{-}M
=
\frac{1}{2}
\sum_{c=0}^{1} M_c,
\label{eq:macro_metric}
\end{equation}
where $M_c$ denotes the corresponding class-wise metric for class $c$. The receiver operating characteristic curve is obtained by varying the decision threshold and plotting the true-positive rate against the false-positive rate.
%
% \begin{equation}
% \mathrm{TPR}
% =
% \frac{\mathrm{TP}}
% {\mathrm{TP}+\mathrm{FN}},
% \qquad
% \mathrm{FPR}
% =
% \frac{\mathrm{FP}}
% {\mathrm{FP}+\mathrm{TN}}.
% \label{eq:roc_rates}
% \end{equation}
% %
The ROC AUC summarizes this curve as
\begin{equation}
\mathrm{ROC\ AUC}
=
\int_{0}^{1}
\mathrm{TPR}(\mathrm{FPR})\,
d(\mathrm{FPR}).
\label{eq:roc_auc}
\end{equation}

\subsection{Dataset}
\label{subsec:dataset}

The experiments use an augmented version of the Electrical Grid Stability Simulated Dataset from the UCI Machine Learning Repository~\cite{misc_electrical_grid_stability_simulated_data__471}. The data set has been used in previous studies on power system stability and adversarial manipulation~\cite{mulo2023towards,EFATINASAB2025101799,alaerjan2025online,EFATINASAB2025101662} and is, to our knowledge, the only publicly available data set derived from the DSGC system. It contains 60,000 simulated configurations of a four-node DSGC system composed of one producer and three consumers, including 21,720 stable and 38,280 unstable configurations. Each configuration is described by the twelve input variables discussed in Section~\ref{sec:system_model}. The data set contains no attack samples; all FDI scenarios used in our experiments are generated separately and are excluded from model training.

\subsection{Attack Impact Evaluations}
\label{subsec:attack_impact}

To examine whether the considered FDI attacks can meaningfully alter a conventional AI-based stability decision, we train an auxiliary supervised multilayer perceptron (MLP) to classify clean DSGC configurations as stable or unstable. MLP-based models have been widely used for data-driven power system stability prediction
~\cite{sym15020289,ALLAL2024108304}. We therefore use the MLP as a representative neural classifier for quantifying attack impact. It is used only for this auxiliary evaluation and is not part of the proposed StarGNN framework.

Specifically, we apply each FDI attack to configurations that the MLP initially classifies correctly and measure how often the manipulated input changes the prediction from stable to unstable (S$\rightarrow$U) which represents a false alarm, in which a stable operating condition is incorrectly classified as unstable or from unstable to stable (U$\rightarrow$S) represents concealment, in which an actually unstable configuration is incorrectly reported as stable. 

The model contains three hidden layers with 128, 64, and 32 neurons and uses GELU activation, layer normalization, and dropout. 
The attacks were applied separately to held-out stable and unstable configurations.

The MLP also outputs a predicted probability for the unstable class. A value close to zero indicates that the model considers the configuration stable, whereas a value close to one indicates a strong unstable prediction. Before the attacks were applied, the mean predicted instability probability was approximately $0.0297$ for stable samples and $0.9717$ for unstable samples. On the exact StarGNN attack-source set, the clean model correctly classified $98.96\%$ of the unstable configurations. Table~\ref{tab:bidirectional_attack_impact} summarizes the effect of the attacks on the auxiliary supervised stability predictor.

\begin{table}[t]
\centering
\footnotesize
\caption{Impact of the evaluation attacks on the auxiliary supervised stability predictor. S$\rightarrow$U denotes the percentage of initially correct stable predictions changed to unstable, while U$\rightarrow$S denotes the percentage of initially correct unstable predictions changed to stable. $M_S$ and $M_U$ denote the mean predicted probability of the unstable class after attacks on stable and unstable configurations, respectively.}
\label{tab:bidirectional_attack_impact}
\renewcommand{\arraystretch}{1.08}

%\resizebox{\columnwidth}{!}{%
\begin{tabular}{lcccc}
\toprule
Attack
& S$\rightarrow$U (\%)
& U$\rightarrow$S (\%)
& $M_S$
& $M_U$
\tabularnewline
\midrule

A1
& 16.73
& 45.10
& 0.1648
& 0.5305
\tabularnewline

A2
& \textbf{27.65}
& 18.40
& 0.2684
& 0.8007
\tabularnewline

A3
& 1.73
& 1.70
& 0.0344
& 0.9518
\tabularnewline

A4
& 0.69
& 0.23
& 0.0326
& 0.9701
\tabularnewline

A5
& 20.66
& 39.57
& 0.2025
& 0.5875
\tabularnewline

A6
& 22.29
& \textbf{48.43}
& 0.2182
& 0.4991
\tabularnewline

A7
& 1.04
& 0.56
& 0.0327
& 0.9676
\tabularnewline

A8
& 0.66
& 0.28
& 0.0324
& 0.9696
\tabularnewline

A9
& 17.74
& 24.91
& 0.1753
& 0.7280
\tabularnewline

A10
& N/A
& 39.73
& N/A
& 0.5871
\tabularnewline

\bottomrule
\end{tabular}%
%}%resizebox

\end{table}

\paragraph{False-alarm and concealment effects.}
Table~\ref{tab:bidirectional_attack_impact} shows that the effectiveness of an attack depends strongly on its objective and the manipulated feature group. Coordinated FDI (A6) produces the strongest non-adaptive concealment effect, changing $48.43\%$ of initially correct unstable predictions to stable and reducing the mean instability probability from approximately $0.9717$ to $0.4991$. Reaction-time bias (A1) produces a similar effect, with a $45.10\%$ unstable-to-stable flip rate despite modifying only reaction-time variables. Node takeover (A5) and replay style substitution (A9) also produce substantial concealment, with flip rates of $39.57\%$ and $24.91\%$, respectively.
The behavior is different for the false alarm objective. Price-response bias (A2) produces the largest stable-to-unstable flip rate at $27.65\%$, while its concealment rate is lower at $18.40\%$. Reaction-time bias (A1) shows the opposite tendency, being considerably more effective at concealing unstable configurations than at creating false alarms. This indicates that different DSGC feature groups affect the stability decision in different directions.

\paragraph{Limited impact of power-based manipulations.}
Attacks that primarily modify the power profile have a much smaller effect on the auxiliary classifier. Power imbalance (A3) produces flip rates of only $1.73\%$ and $1.70\%$ in the two directions, while balance-preserving power (A4), load redistribution (A7), and local load redistribution (A8) remain close to or below $1\%$. Thus, modifying more reported values does not necessarily lead to a larger change in the predicted stability state. Under the considered setting, the choice of manipulated feature appears more important than the number of modified entries.

\paragraph{Adaptive single-node concealment.}
The adaptive attack (A10) is evaluated only in the unstable-to-stable direction because it is designed specifically to conceal unstable operating points. It changes $39.73\%$ of initially correct unstable MLP predictions to stable and reduces the mean instability probability to $0.5871$. This is a substantial effect despite the attacker controlling only a single participant. The result provides an additional stress test showing that the manipulated reports found through adaptive search can also meaningfully affect a conventional supervised stability classifier.

The auxiliary MLP confirms that several of the evaluation attacks are capable of producing meaningful changes in a supervised stability decision. The strongest attack depends on the adversarial objective as price-response bias is most effective at inducing false alarms, whereas coordinated FDI and reaction-time bias produce the strongest non-adaptive concealment effects. These results provide an attack-impact reference for the StarGNN evaluation in Section~\ref{subsec:attack_results}, where the main question is whether the proposed stable region approach continues to identify such manipulated configurations as abnormal.

% \subsection{Proposed Stability Prediction Results}
\subsection{StarGNN Stability Prediction Evaluation} \label{subsec:stability_results}

First, we investigated StarGNN's ability to detect instability in our dataset.

Our system is trained, and the threshold is calibrated without using unstable configurations. The model is trained only on stable samples and their physics-constrained pseudo-negative counterparts.

The stability prediction results are reported using the threshold obtained from the $0.90$ quantile of the stable validation scores (threshold value is discussed in the ablation studies in Section~\ref{subsec:ablation}).

StarGNN achieves an accuracy of 0.989, with macro-averaged precision, recall, and F1-score of 0.991, 0.952, and 0.970, respectively. The ROC AUC reaches $0.999$.

The macro-averaged scores are important because the test set contains substantially more unstable than stable configurations. The macro F1-score of 0.970 indicates that the detector performs well across both classes.
The ROC AUC of $0.999$ shows that the continuous risk score provides a very strong ordering of stable and unstable configurations across different possible thresholds. In addition to the full ROC AUC, we report the standardized partial ROC AUC over the low-false-alarm region $\mathrm{FPR}\leq0.05$ which was $0.993$. This measure focuses the comparison on the operating range most relevant to practical monitoring, where high detection sensitivity is required without causing excessive false alarms.

Moreover, StarGNN's performance on the unseen unstable set suggests that the pseudo-negatives used during training provided a useful contrastive signal to shape the boundary around the stable operating region.

\begin{figure}[t]
    \centering
    \includegraphics[width=0.75\columnwidth]{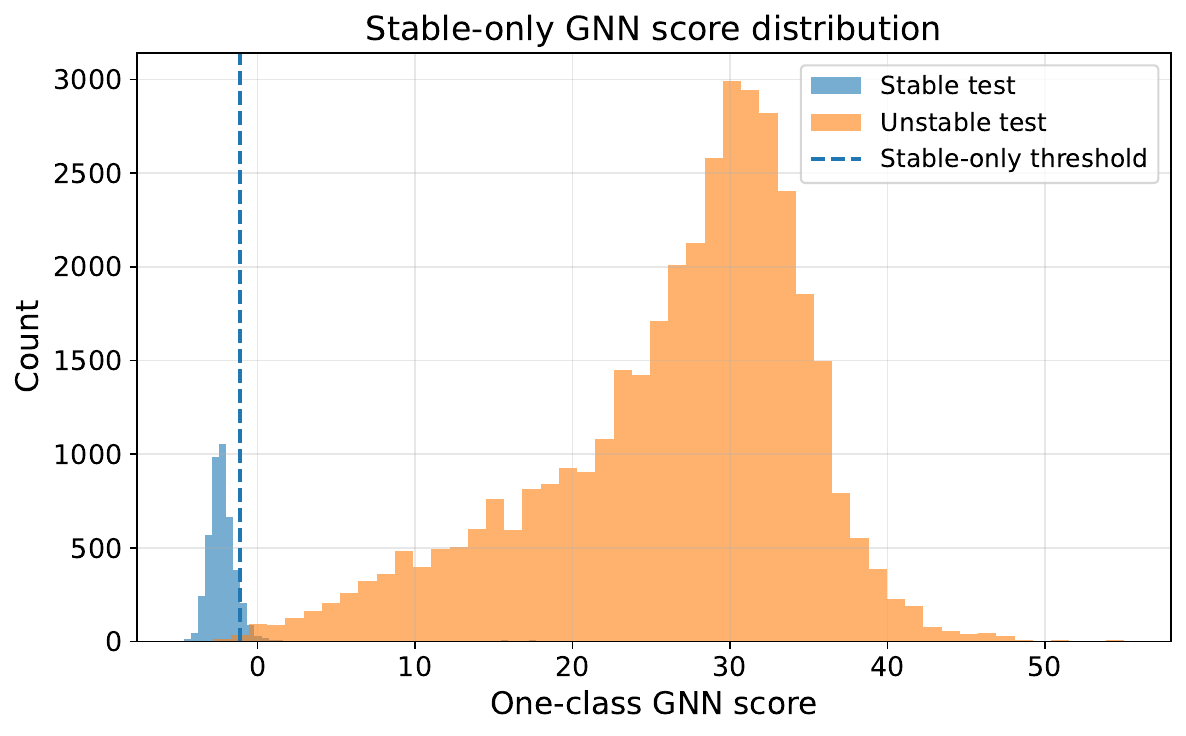}
    \caption{Distribution of the raw StarGNN risk scores for held-out stable and unstable DSGC configurations.
    % The dashed line indicates the threshold calibrated at the $0.90$ quantile of the stable validation scores.
    }
    \label{fig:stability_score_distribution}
\end{figure}
Figure~\ref{fig:stability_score_distribution} provides a clearer view of the separation produced by the learned risk score. The stable test samples are concentrated within a relatively narrow low score region, whereas the unstable configurations are distributed mainly at considerably higher scores. Most samples, therefore, lie on the expected side of the stable-only threshold.
Furthermore, to examine whether the generated pseudo-negatives provide a consistent training direction, we generated fresh perturbations from held-out stable test configurations and compared each pseudo-negative score with the score of its clean source sample. As shown in Appendix Fig.~\ref{fig:paired_score_gap}, $96.55\%$ of the clean--pseudo-negative pairs produced a positive score difference, while $86.54\%$ satisfied the full ranking margin of $1.0$. The median score increase was $7.19$. A one-sided Wilcoxon signed-rank test also confirmed that the pseudo-negative scores were significantly higher than their paired clean scores ($p<10^{-16}$).
These results show that the generator produces controlled deviations that the trained model consistently places outside the low risk region associated with stable operation. This confirms that they provide informative directional stress samples for learning the boundary around the observed stable region.

\subsection{Unseen FDI Attack Detection Results}
\label{subsec:attack_results}

The attack evaluation uses the same StarGNN risk score and decision threshold as for stability prediction. The threshold is calibrated once at the $0.90$ quantile of the stable validation scores and is not adjusted for any attack family. We evaluate the attacks under the two objectives defined in the threat model. For the \textit{false-alarm objective}, attacks are applied to clean stable configurations, and the detector must distinguish the manipulated reports from normal operation. For the \textit{concealment objective}, attacks are applied to unstable configurations, and the relevant question is whether the manipulated input can be moved into the learned stable region.
\begin{table}[t]
\centering
\small
\caption{Stable-origin unseen-FDI detection results at $q=0.90$. The attacks are applied to clean stable configurations and represent the false-alarm objective.}
\label{tab:stable_origin_attack_results}
\renewcommand{\arraystretch}{1.0}
\footnotesize

\begin{tabular}{lcccccc}
\toprule
Attack
& Acc.
& M Prec.
& M Rec.
& M F1
& Attack Rec.
& ROC AUC
\tabularnewline
\midrule

A1
& 0.913
& 0.913
& 0.913
& 0.913
& 0.921
& 0.966
\tabularnewline

A2
& 0.939
& 0.941
& 0.939
& 0.939
& 0.973
& 0.985
\tabularnewline

A3
& 0.842
& 0.848
& 0.842
& 0.842
& 0.780
& 0.906
\tabularnewline

A4
& 0.886
& 0.886
& 0.886
& 0.886
& 0.867
& 0.944
\tabularnewline

A5
& 0.915
& 0.915
& 0.915
& 0.915
& 0.926
& 0.969
\tabularnewline

A6
& 0.938
& 0.940
& 0.938
& 0.938
& 0.971
& 0.985
\tabularnewline

A7
& 0.920
& 0.920
& 0.920
& 0.920
& 0.935
& 0.969
\tabularnewline

A8
& 0.849
& 0.853
& 0.849
& 0.848
& 0.793
& 0.912
\tabularnewline

A9
& 0.904
& 0.904
& 0.904
& 0.904
& 0.904
& 0.960
\tabularnewline

\midrule
\textbf{Mean}
& \textbf{0.900}
& \textbf{0.902}
& \textbf{0.900}
& \textbf{0.900}
& \textbf{0.896}
& \textbf{0.955}
\tabularnewline

\bottomrule
\end{tabular}

\vspace{1mm}
{\scriptsize
M Prec., M Rec., and M F1 denote macro precision, macro recall, and macro F1-score.
Attack Rec. is the proportion of attacked stable configurations classified as abnormal.
}

\end{table}
\begin{table}[t]
\centering
\footnotesize
\caption{Post-attack instability detection at $q=0.90$ for attacks applied to unstable configurations. This evaluation represents the instability-concealment objective.}
\label{tab:unstable_origin_attack_results}
\renewcommand{\arraystretch}{1.0}

\begin{tabular}{lccccccc}
\toprule
Attack
& Acc.
& M Prec.
& M Rec.
& M F1
& U Rec.
& $\Delta$Rec.
& ROC AUC
\tabularnewline
\midrule

A1
& 0.938
& 0.940
& 0.938
& 0.938
& 0.972
& 0.027
& 0.987
\tabularnewline

A2
& 0.950
& 0.954
& 0.950
& 0.950
& 0.996
& 0.003
& 0.997
\tabularnewline

A3
& 0.952
& 0.956
& 0.952
& 0.951
& 0.998
& 0.000
& 0.998
\tabularnewline

A4
& 0.951
& 0.955
& 0.951
& 0.951
& 0.998
& 0.001
& 0.999
\tabularnewline

A5
& 0.945
& 0.948
& 0.945
& 0.945
& 0.985
& 0.013
& 0.992
\tabularnewline

A6
& 0.943
& 0.945
& 0.943
& 0.942
& 0.980
& 0.018
& 0.990
\tabularnewline

A7
& 0.952
& 0.956
& 0.952
& 0.952
& 0.999
& 0.000
& 0.999
\tabularnewline

A8
& 0.952
& 0.956
& 0.952
& 0.951
& 0.998
& 0.000
& 0.999
\tabularnewline

A9
& 0.946
& 0.949
& 0.946
& 0.946
& 0.987
& 0.011
& 0.994
\tabularnewline

A10
& 0.902
& 0.902
& 0.902
& 0.902
& 0.899
& 0.100
& 0.975
\tabularnewline

\midrule
\textbf{Mean}
& \textbf{0.943}
& \textbf{0.946}
& \textbf{0.943}
& \textbf{0.942}
& \textbf{0.981}
& \textbf{0.017}
& \textbf{0.993}
\tabularnewline

\bottomrule
\end{tabular}

\vspace{1mm}
{\scriptsize
U Rec. denotes the recall of attacked unstable configurations. $\Delta$Rec.
denotes the reduction relative to the clean unstable recall of 0.9995.
M Prec., M Rec., and M F1 denote macro precision, macro recall, and macro F1-score.
For the adaptive attack A10, we used 25 additional binary-alarm queries.
}

\end{table}

\paragraph{Detection under the false alarm induction objective}
Table~\ref{tab:stable_origin_attack_results} shows that StarGNN detects most manipulated stable configurations despite never observing these attacks during training. Price response bias and coordinated FDI are the most readily detected cases, with attack recalls of $0.973$ and $0.971$, respectively. Reaction time bias, node takeover, load redistribution, and replay node substitution are also detected reliably. The replay result (A9) is particularly relevant because the substituted values originate from another valid stable configuration, yet the resulting participant level combination is still frequently recognized as inconsistent with the learned stable operating region.

The more difficult cases are those that make relatively limited changes to the power profile. Power imbalance (A3) and local load redistribution (A8) reach attack recalls of $0.780$ and $0.793$, respectively, while the balance-preserving power attack (A4) reaches $0.867$. These are also among the attacks with the smallest effect on the auxiliary supervised MLP used in Section~\ref{subsec:attack_impact} to quantify attack impact.  
The attacks that remain closest to the learned stable region are therefore also the least effective at changing a conventional supervised stability decision. This suggests that the lower detection rates are associated with comparatively mild manipulations rather than with attacks that strongly alter the predicted system state while remaining undetected.

\paragraph{Resistance to instability-concealment attacks.}
Table~\ref{tab:unstable_origin_attack_results} considers the more critical setting in which the source configuration is genuinely unstable. Before manipulation, StarGNN identifies $99.9\%$ of the selected unstable configurations. After the nine non-adaptive attacks (A1-A9), unstable recall remains above $97\%$ in all cases, indicating that these manipulations generally fail to move an unsafe configuration into the stable region learned by StarGNN.

This result is particularly relevant when considered together with the attack impact reported in Table~\ref{tab:bidirectional_attack_impact}. 
Coordinated FDI (A6), reaction time bias (A1), node takeover (A5), and replay style node substitution (A9) cause unstable to stable flips in the MLP in $48.43\%$, $45.10\%$, $39.57\%$, and $24.91\%$ of the corresponding cases, respectively. Within the same attack families, StarGNN retains unstable recall values of $0.980$, $0.972$, $0.985$, and $0.987$. Thus, attacks that can substantially alter a conventional supervised classification still leave most manipulated unstable configurations outside the stable operating region learned by StarGNN.
Reaction time bias (A1) produces the largest reduction in unstable recall among the non-adaptive attacks, from $0.999$ on clean unstable configurations to $0.972$ after manipulation. Coordinated FDI (A6) follows at $0.980$, while node takeover (A5) and replay style substitution (A9) retain recalls of $0.985$ and $0.987$, respectively. The remaining attacks have only a small effect on instability detection. Price response bias (A2) retains an unstable recall of $0.996$, while power imbalance (A3), balance-preserving power (A4), load redistribution (A7), and local load redistribution (A8) all remain above $0.998$.

The contrast between the two attacker objectives is also notable. Manipulations of otherwise stable configurations are harder to detect than attempts to conceal already unstable configurations. In the latter case, the source configuration already lies outside the learned stable region, and most non-adaptive manipulations are insufficient to move it back into that region. This is reflected in the consistently high post-attack unstable recall, compared with the wider variation observed for stable-origin attack detection.

\paragraph{Impact of adaptive binary feedback.}

The single node adaptive concealment attack (A10) provides a more demanding test. With 25 additional binary alarm queries, unstable recall decreases to $0.899$, corresponding to a reduction of $0.100$ from the clean value. The ROC AUC remains $0.975$, indicating that the continuous StarGNN risk score still provides substantial separation between clean, stable, and adaptively manipulated, unstable configurations.
The auxiliary MLP provides a complementary measure of attack impact. Under the same adaptive manipulation, $39.73\%$ of its initially correct unstable predictions flip to the stable class, confirming that the attack can substantially affect a conventional supervised stability classifier. Against StarGNN, the detector evasion rate is $10.06\%$. 
The MLP result establishes that the manipulation can change a supervised stability decision, whereas the StarGNN result assesses whether the proposed stable region approach continues to recognize the manipulated configuration as abnormal. Even with adaptive binary feedback, most attacked unstable configurations remain outside the stable region learned by StarGNN.

The adaptive setting should be interpreted as a stress test rather than as an attacker's capability that is always available in practice. It assumes that the adversary can repeatedly submit modified reports, observe the resulting binary alarm, and associate each response with the corresponding candidate. Such feedback may be delayed or unavailable in deployments where alarms are visible only to system operators. Repeated probing may also create a detectable sequence of related submissions. Stateful defenses for query-based black-box attacks have shown that query histories can be used to identify iterative or highly similar inputs~\cite{10.1145/3385003.3410925,li2022blacklight}, while rate limiting would further restrict the number of opportunities available for adaptive search.

None of the evaluated attacks is used during training, and all attack samples are generated only after the StarGNN model and its decision threshold have been fixed. The physics-constrained pseudo-negative generator modifies the reaction time and price response variables, whereas the evaluation also includes power manipulation, node compromise, load redistribution, replay-style substitution, coordinated FDI, and adaptive binary feedback search. The observed detection performance, therefore, cannot be explained simply by reproducing the perturbation patterns used during training.

\section{Discussion}\label{sec:discussion}
In this section, we start by ablation studies presented in Section~\ref{subsec:ablation}. Section~\ref{sec:baseline_comparison} compares StarGNN with baselines and Section~\ref{sec:computational_complexity} discusses computational complexity. 

\subsection{Ablation Study}
\label{subsec:ablation}

We first examine the effect of the threshold quantile and then study the main components of the proposed framework.

\paragraph{Threshold Sensitivity}

The decision threshold is obtained from the empirical distribution of the stable validation scores. A higher quantile produces a more permissive threshold, reducing false alarms on stable configurations at the cost of lower sensitivity to unstable or manipulated samples near the learned
boundary. The complete sensitivity analysis for clean stability prediction, the six generic attacks, the three literature-inspired attacks, and the adaptive attack is reported in Appendix Fig.~\ref{fig:threshold_ablation}.

The fixed unseen attacks remain detectable across most of the examined threshold range, whereas the adaptive attack is more sensitive to increasingly permissive thresholds. At the selected operating point of $\alpha=0.90$, clean stability prediction achieves a macro F1-score of $0.970$. The generic
attacks obtain a mean attack recall of $0.988$, with a worst-case recall of $0.972$, while the literature-inspired attacks achieve a mean recall of $0.995$ and a worst-case recall of $0.987$. The threshold-specific adaptive attack yields a macro F1-score of $0.902$ and an attack recall of $0.899$.
These results indicate that $\alpha=0.90$ provides a reasonable balance between reducing false alarms on normal configurations and retaining sensitivity to both fixed unseen attacks and the stronger binary feedback adaptive attack.

\paragraph{Contribution of the Main Components}

We next remove or replace individual components while keeping the data partitions, optimization settings, and threshold quantile unchanged. Table~\ref{tab:component_ablation} reports the main results on the genuine stability prediction task.
The non-topological baseline is a parameter-matched MLP that receives the same $4\times3$ node tensor after flattening it into a twelve dimensional vector. It therefore has access to the same input values, but it does not perform producer--consumer message passing. In the clean-only variant, pseudo-negatives are removed, and the model is trained using only the SVDD compactness and variance terms. Since its risk head receives no supervision, the distance from the stable center is used as its test score.

\begin{table}[t]
\centering
\caption{Component ablation at $q=0.90$.}
\label{tab:component_ablation}
\footnotesize
\setlength{\tabcolsep}{2.2pt}
\renewcommand{\arraystretch}{1.08}
\begin{tabular}{lccccc}
\toprule
Configuration
& Acc.
& \shortstack{M\\Prec.}
& \shortstack{M\\Rec.}
& \shortstack{M\\F1}
& \shortstack{ROC\\AUC} \\
\midrule
\textbf{Full framework}
& \textbf{0.989}
& \textbf{0.991}
& \textbf{0.952}
& \textbf{0.970}
& \textbf{0.999} \\

Without ranking
& 0.987
& 0.982
& 0.948
& 0.964
& 0.998 \\

MLP, no topology
& 0.969
& 0.903
& 0.940
& 0.920
& 0.989 \\

Without pseudo-negatives
& 0.101
& 0.051
& 0.500
& 0.092
& 0.494\\
\bottomrule
\end{tabular}
\end{table}
Removing the pseudo-negatives causes the largest drop. The macro F1-score falls from $0.970$ to $0.092$, and the ROC AUC decreases from $0.999$ to $0.494$. In this setting, training only on stable samples does not provide enough information to establish which departures should receive higher risk scores. The model consequently fails to distinguish the unseen unstable configurations from the stable region. This result shows that the pseudo-negatives are not a minor data augmentation step; they provide the contrast required to learn a useful boundary without using genuine unstable labels.

Replacing StarGNN with the parameter-matched MLP also reduces performance. The macro F1-score falls to $0.920$, despite both models receiving the same twelve standardized features and using the same pseudo-negative samples and loss terms. The difference supports the use of the producer--consumer structure rather than treating the participant variables as an unstructured vector. 

Removing the ranking term produces a smaller, but consistent, decline. The macro F1-score decreases to $0.964$, while ROC AUC falls to $0.998$. Binary separation already distinguishes stable samples from pseudo-negatives, but the pairwise ranking term further encourages every selected pseudo-negative to receive a higher score than the stable configuration from which it was generated.
The ablation results identify the pseudo-negative generator as the main source of contrastive boundary information and the topology-aware architecture as an important part of the final performance. 

\subsection{Comparison with Stable-Only Baselines}
\label{sec:baseline_comparison}

The proposed model was compared with established anomaly detection baselines such as One-Class SVM, Isolation Forest , PCA reconstruction, an MLP autoencoder, and Deep SVDD.
All methods were trained using the same clean stable training partition, and the feature scaler was fitted only on this partition. Genuine unstable configurations and attacked observations were excluded from baseline training and threshold calibration. The conventional baselines were retained in their standard form and were not provided with the physics-constrained pseudo-negatives used to train StarGNN.

The comparison is based on ROC AUC because it evaluates the complete ranking of the anomaly scores independently of a particular operating threshold. This is important because the methods produce scores on different numerical scales and because the complete stability test set is strongly imbalanced. 
% ROC AUC measures how consistently a detector assigns a higher abnormality score to an unstable or manipulated configuration than to a clean stable configuration.
\begin{table}[t]
\centering
\footnotesize
\caption{Threshold-independent comparison of the proposed method and clean stable baselines. Attack ROC AUC values are averaged across the nine
non-adaptive evaluation attacks.}
\label{tab:baseline_auc}
\setlength{\tabcolsep}{4pt}
\renewcommand{\arraystretch}{1.08}

\begin{tabular}{lccc}
\toprule
\textbf{Method}
& \shortstack{\textbf{Stability} \\ \textbf{AUC}}
& \shortstack{\textbf{False-alarm} \\ \textbf{AUC}}
& \shortstack{\textbf{Concealment} \\ \textbf{AUC}}
\\
\midrule

PCA reconstruction
& 0.786
& 0.657
& 0.817
\\

MLP autoencoder
& 0.755
& 0.798
& 0.886
\\

Deep SVDD
& 0.725
& 0.797
& 0.783
\\

Isolation Forest
& 0.659
& 0.721
& 0.811
\\

One-Class SVM
& 0.541
& 0.811
& 0.825
\\

\textbf{StarGNN (proposed)}
& \textbf{0.999}
& \textbf{0.955}
& \textbf{0.993}
\\

\bottomrule
\end{tabular}

\vspace{1mm}
{\scriptsize
Stability AUC compares clean stable and unstable configurations.
False-alarm AUC compares clean stable configurations with attacked stable configurations. Concealment AUC compares clean stable configurations with attacked unstable configurations.}

\end{table}

Three evaluations are reported in Table~\ref{tab:baseline_auc}. 

StarGNN achieves a ROC AUC of $0.999$ for genuine instability, $0.955$ for stable-origin attack detection, and $0.993$ for attacked unstable configurations. It therefore maintains a strong score ordering across all three settings. 

The difference between the two attack columns is also interesting as most conventional baselines achieve higher ROC AUC in the unstable-origin setting than in the stable-origin setting. For example, the MLP autoencoder increases from $0.798$ to $0.886$, while PCA reconstruction increases from $0.657$ to $0.817$. This does not necessarily indicate stronger identification of the injected manipulation. The underlying inputs are already unstable, and their physical abnormality may remain visible even after the attack. 
One-Class SVM shows a different limitation. It reaches a stable-origin attack ROC AUC of $0.811$ despite obtaining only $0.541$ for stability prediction task. 

Similarly, reconstruction-based methods retain some anomaly ranking ability, but their results remain below those of StarGNN, particularly for genuine instability and subtle stable-origin manipulations.

\subsection{Computational Complexity}
\label{sec:computational_complexity}

Let $n$ denote the number of stable training configurations, $K$ the number of precomputed stress variants per configuration, $V$ the number of graph nodes, $H$ the hidden dimension, and $L$ the number of message-passing layers. Constructing the bounded random stress bank requires $\mathcal{O}(nKD)$ operations and memory, where $D$ is the number of input variables. In the present implementation, $K=16$ and $D=12$ are fixed; therefore, stress-bank generation and storage scale linearly with the number of training configurations, i.e., $\mathcal{O}(n)$. Although all $K$ variants are retained, only one stress configuration is sampled for each stable observation in a training iteration, so the training cost does not increase by a factor of $K$.

For one configuration, the node encoder and linear transformations in the message-passing layers require approximately $\mathcal{O}(LVH^{2})$ operations. Message aggregation over the star topology is linear in the number of nodes, $\mathcal{O}(LVH)$, since the producer receives an aggregate of the consumer messages and each consumer receives the producer message; no all pairs node interaction is computed. Because training evaluates both the clean and selected stressed configurations, one epoch has complexity $\mathcal{O}(nLVH^{2})$, up to a constant factor of two. Recomputing the SVDD center every $R$ epochs introduces an additional $\mathcal{O}((E/R)nLVH^{2})$ cost over $E$ training epochs. Hence, the complete training procedure has complexity $\mathcal{O}\!\left(EnLVH^{2}+\frac{E}{R}nLVH^{2}\right)$ which is linear in $n$ when $E$, $R$, $L$, $V$, and $H$ are fixed. In the evaluated model, $E=150$, $R=10$, $L=2$, $V=4$, and $H=64$. 
During inference, each observation is converted into a four-node graph and processed through a single forward pass. Scoring $n_{\mathrm{test}}$ configurations therefore requires $\mathcal{O}(n_{\mathrm{test}}LVH^{2})$ operations. Furthermore, the proposed model required an average of 1.0864 seconds per epoch, with a total training time of 163.1861 seconds. All experiments were conducted on the free Kaggle cloud platform using an Intel Xeon CPU at 2.20 GHz, 32 GB of RAM, and an NVIDIA Tesla T4 GPU under Ubuntu Linux.

\paragraph{Scalability} For a producer--consumer star graph with $V$ nodes, the number of edges is $V-1$. The message-passing cost therefore grows linearly with the number of participants rather than quadratically, since StarGNN does not evaluate all-pairs node interactions. More precisely, the per-configuration forward cost can be written as $\mathcal{O}(LVH^{2}+LVH)$, where the first term accounts for the shared linear transformations and the second for message aggregation. For fixed $L$ and $H$, this gives linear computational growth with $V$.
The model size itself does not grow with the number of participants. The same node encoder and message-passing transformations are reused across nodes, and the consumer representations are combined through permutation invariant
aggregation. Thus, increasing the number of consumers increases the amount of computation required for a forward pass, but does not require a separate set of parameters for each additional participant.

% % \input{Sections/09-Takeaways}
\section{Conclusion}
\label{sec:conclusion}
This paper presented a stable-only framework that uses one model and one threshold for both DSGC stability prediction and unseen attack detection. The model is trained on stable configurations and physics-constrained pseudo-negatives, which provide controlled departures from the stable region without being treated as genuinely unstable samples. StarGNN further preserves the producer--consumer structure instead of processing the twelve inputs as an unstructured vector.
The same risk score performs well on stability prediction and provides acceptable detection across the nine unseen attacks considered in the threat model. The ablation results show that pseudo-negative generation is essential for learning a useful boundary, while the lower performance of the parameter-matched MLP confirms the benefit of topology-aware graph learning.
\paragraph{Limitation and Future Work}
The main limitation concerns the size of the evaluated DSGC system. The benchmark used in this work contains one producer and three consumers and is the only publicly available data set derived from the DSGC system. The reported detection results are therefore necessarily validated in this four-node setting. The StarGNN architecture does not rely on all-pairs message passing, and its shared transformations and consumer aggregation lead
to linear computational growth with the number of participants for fixed model dimensions. This provides a favorable computational scaling property, but it does not establish that the same detection performance will be retained in larger DSGC systems. Larger systems may exhibit operating patterns and attack
interactions that are not represented in the available benchmark. Validation on larger DSGC configurations will therefore require a suitable larger-scale simulation environment, and is foreseen as future work.

\bibliographystyle{ACM-Reference-Format}
\bibliography{bibliography1.bib}
\appendix
\section{Open Science}
We make the code and experimental data publicly available at: \url{https://anonymous.4open.science/r/System-Aware-Graph-Boundary-Learning-8E0E}

\section{StarGNN framework details}
\begin{figure}[tbh]
    \centering
    \includegraphics[width=0.70\columnwidth]{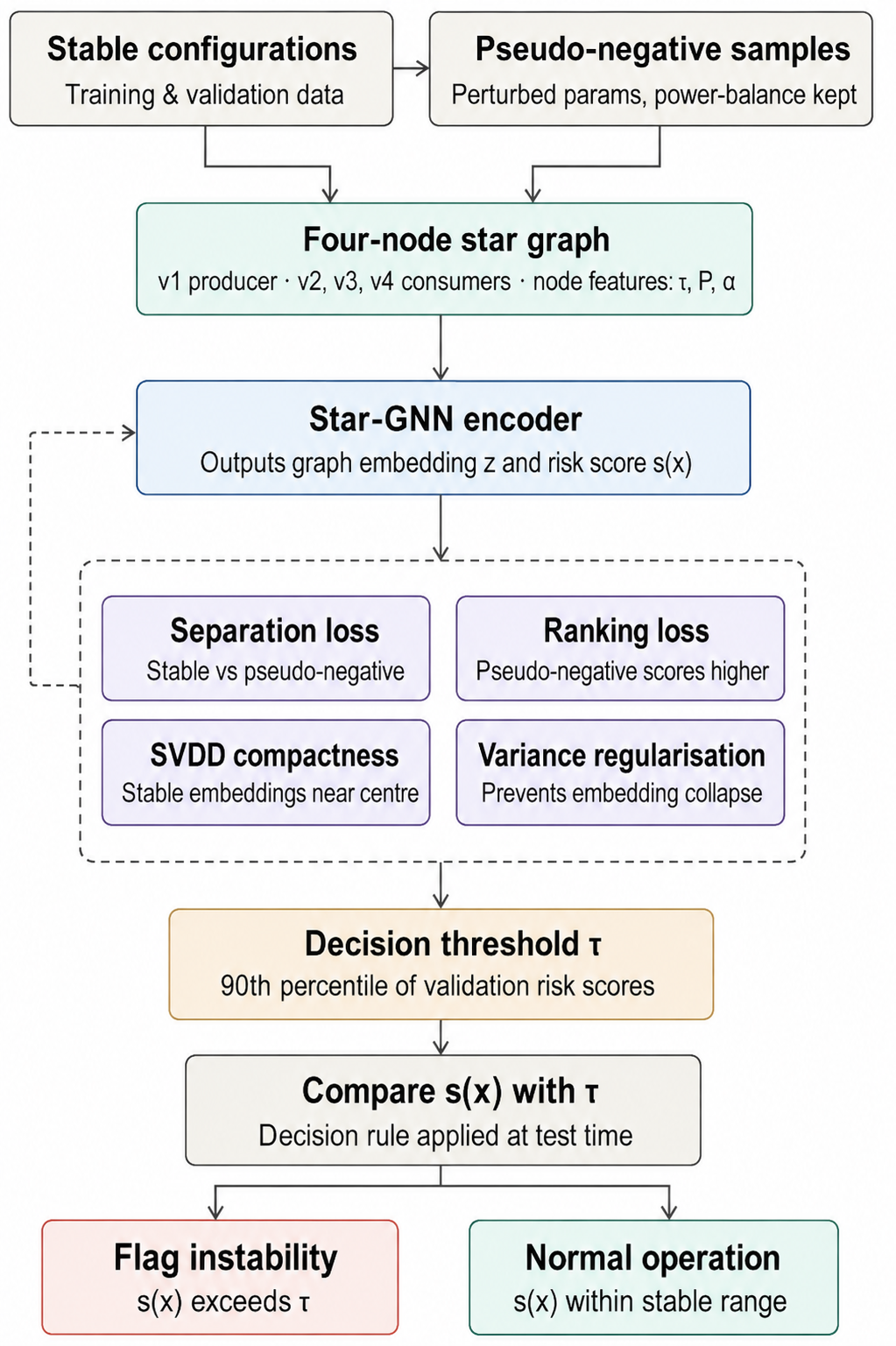}
    \caption{Overview of the proposed framework.}
    \label{fig:framework}
\end{figure}

More details regarding the configuration of StarGNN are illustrated in Figure~\ref{fig:framework}. Stable configurations and pseudo-negative samples, all representing four-node star graphs, are used to train the StarGNN encoder. It employs separation and ranking loss, SVDD compactness, and regularization to learn the shape of the stable data, as discussed in Section~\ref{sec:methodology}. Then, a validation set from the same stable and pseudo-negative sample dataset is used to determine a threshold for the testing phase, where the model is asked to flag anomalous behaviors. 

\section{Detailed FDI Attack Construction}
\label{app:attack_details}

This appendix provides the exact construction of the FDI attacks summarized in
Table~\ref{tab:attack_summary}. All attacks are generated only after the
StarGNN model and its decision threshold have been fixed. No attack samples are
used during model training or threshold calibration.

Let $R_k$ denote the range of feature $k$ computed from the clean training
data. These ranges are used to scale the perturbation magnitudes. After each
manipulation, the affected features are clipped to their predefined admissible
bounds to avoid invalid reported values.

\subsection{Non-Adaptive FDI Attacks}

\paragraph{Reaction-time bias attack.}
The attacker selects between one and four reaction-time variables and modifies
each selected $\tau_i$ according to
\begin{equation}
    \Delta \tau_i
    =
    s_i \rho_i R_{\tau_i},
    \qquad
    \rho_i \sim \mathcal{U}(0.15,0.50),
\end{equation}
where $s_i\in\{-1,+1\}$ determines the direction of the perturbation. The
manipulated value is therefore
\begin{equation}
    \widetilde{\tau}_i
    =
    \tau_i+\Delta\tau_i.
\end{equation}
This attack represents falsification of the reported response-time behavior of
DSGC participants. It adapts the idea of time-delay attacks studied in
power-grid control systems~\cite{10.1145/3055386.3055392}. Since the DSGC
dataset does not contain packet-level transmission delays, the attack is
implemented through the reported reaction-time parameter.

\paragraph{Price-response bias attack.}
The attacker selects between one and four price-response coefficients and
modifies each selected $g_i$ as
\begin{equation}
    \Delta g_i
    =
    s_i \rho_i R_{g_i},
    \qquad
    \rho_i \sim \mathcal{U}(0.15,0.55),
\end{equation}
with $s_i\in\{-1,+1\}$. The resulting value is
\begin{equation}
    \widetilde{g}_i
    =
    g_i+\Delta g_i.
\end{equation}
The attack represents falsification of the reported participant response to
frequency-dependent prices and changes how the reported economic-control
behavior is presented to the monitoring system.

\paragraph{Power-imbalance attack.}
The attacker modifies between one and three consumer-power values. For each
selected consumer,
\begin{equation}
    \Delta p_i
    =
    s_i \rho_i R_{p_i},
    \qquad
    \rho_i \sim \mathcal{U}(0.15,0.50),
\end{equation}
where $s_i\in\{-1,+1\}$. The producer power report $p_1$ is left unchanged.
The attack is related to load-altering attacks in which compromised demand
measurements are modified to affect grid operation~\cite{5976424}. In this
case, the adversary controls consumer side reports but does not coordinate the
changes to preserve producer--consumer power balance.

\paragraph{Balance-preserving power attack.}
The attacker again modifies between one and three consumer-power values, but
uses a smaller perturbation interval,
\begin{equation}
    \Delta p_i
    =
    s_i \rho_i R_{p_i},
    \qquad
    \rho_i \sim \mathcal{U}(0.10,0.40).
\end{equation}
After modifying the selected consumers, the producer power is recomputed as
\begin{equation}
    \widetilde{p}_1
    =
    \left|
    \widetilde{p}_2+
    \widetilde{p}_3+
    \widetilde{p}_4
    \right|.
\end{equation}
The manipulated report therefore remains consistent with the basic
producer--consumer power balance relation. This attack follows the general
idea of constraint-aware or stealthy FDI, where several measurements are
jointly changed so that simple system consistency checks remain satisfied.

\paragraph{Node-takeover attack.}
A single participant $i$ is selected. All three of its reported attributes,
$\tau_i$, $p_i$, and $g_i$, are modified simultaneously. Each feature receives
an independently selected perturbation direction, and the magnitude is sampled
between $20\%$ and $60\%$ of the corresponding training-derived feature range.
All modified values are clipped to their admissible bounds.

This attack models compromise of a participant's reporting unit, allowing the
attacker to falsify the complete local report rather than only one measurement
type.

\paragraph{Coordinated FDI attack.}
The attacker combines reaction time and price response manipulation with either
the power-imbalance attack or the balance-preserving power attack. The attack
therefore alters multiple feature groups within the same reported DSGC
configuration.

This scenario represents a stronger adversary capable of coordinating
falsified reports across control-related and power-related variables rather
than manipulating a single measurement category.

\subsection{Literature-Inspired FDI Scenarios}

\paragraph{Load-redistribution attack.}
Following the load-redistribution principle in~\cite{10521762}, the attacker
changes the distribution of consumer demand while attempting to preserve the
aggregate reported demand.

For a clean configuration, the original aggregate consumer demand is
\begin{equation}
    P_{\mathrm{load}}
    =
    p_2+p_3+p_4.
\end{equation}
The attack generates unequal random redistribution weights
$w_2,w_3,w_4$ satisfying
\begin{equation}
    w_2+w_3+w_4=1,
\end{equation}
and constructs the modified consumer reports as
\begin{equation}
    \widetilde{p}_i
    =
    w_i P_{\mathrm{load}},
    \qquad
    i\in\{2,3,4\}.
\end{equation}
The producer power is then recomputed according to
\begin{equation}
    \widetilde{p}_1
    =
    \left|
    \widetilde{p}_2+
    \widetilde{p}_3+
    \widetilde{p}_4
    \right|.
\end{equation}
Before boundary clipping, the transformation therefore preserves the aggregate
consumer demand while changing its distribution across participants. The
scenario models an attacker that attempts to hide the manipulation at the
aggregate demand level while altering local reports.

\paragraph{Local load-redistribution attack.}
Local load-redistribution attacks assume that the adversary can manipulate only
a limited part of the system~\cite{6805238}. In our implementation, one or two
consumer-power reports are selected and modified. Another consumer report is
then adjusted in the opposite direction so that the total consumer demand
remains approximately unchanged.

After the local redistribution, the producer power is recomputed as
\begin{equation}
    \widetilde{p}_1
    =
    \left|
    \widetilde{p}_2+
    \widetilde{p}_3+
    \widetilde{p}_4
    \right|.
\end{equation}
This attack models a spatially limited compromise in which the attacker controls
only a subset of consumer measurements rather than all reported loads.

\paragraph{Replay style node substitution.}
A conventional replay attacker records valid measurements and later presents
them again to the monitoring system~\cite{5394956}. The DSGC dataset contains
static operating configurations rather than temporal measurement sequences, so
a direct temporal replay cannot be implemented.

We therefore use a static replay style adaptation. One participant $i$ is
selected, and its complete reported attribute vector
\begin{equation}
    (\tau_i,p_i,g_i)
\end{equation}
is replaced with the corresponding participant attributes taken from another
clean stable configuration. The remaining participants retain their original
values.

The resulting input is therefore assembled from values that are individually
valid and were observed in clean operation, although the substituted
participant may no longer be consistent with the rest of the current
configuration.

\subsection{Adaptive Single Node Concealment Attack}

The adaptive attack targets an unstable operating point $\mathbf{x}$ and seeks
a manipulated report that does not trigger the StarGNN abnormality alarm. For
each attacked sample, one participant $i$ is selected and only its
reaction time and price-response reports,
\begin{equation}
    \mathcal{I}_i=\{\tau_i,g_i\},
\end{equation}
are mutable. All power reports and all attributes belonging to the other
participants remain unchanged.

For each mutable feature $k\in\mathcal{I}_i$, the maximum allowed change is
defined as
\begin{equation}
    |x'_k-x_k|
    \leq
    \epsilon R_k,
\end{equation}
where $R_k$ is the corresponding training-derived feature range. In the
reported experiments,
\begin{equation}
    \epsilon=0.5,
\end{equation}
so each mutable feature may change by at most $50\%$ of its training-derived
range. Candidate values are additionally clipped to the admissible feature
bounds.

The attacker has no access to the StarGNN architecture, learned parameters,
continuous risk score, gradients, or decision threshold. Its only feedback is
the binary alarm returned for each submitted candidate.

Starting from an initial random perturbation, the attacker performs a
limited budget randomized search. Candidate generation combines two forms of
exploration:

\begin{itemize}
    \item \textit{Local exploration}, which samples new perturbations around
    the currently retained candidate; and
    \item \textit{Global exploration}, which samples candidates over the
    complete admissible perturbation region.
\end{itemize}

Each candidate is submitted to StarGNN and the attacker observes only whether
the abnormality alarm is triggered. The attacker is allowed 25 additional
adaptive queries after evaluating the initial random candidate, resulting in
at most 26 detector evaluations per attacked sample.

Among candidates that do not trigger the StarGNN alarm, the attacker retains
the candidate with the largest normalized displacement from the original
configuration. The displacement is measured as
\begin{equation}
    S(\mathbf{x}')
    =
    \left\|
    \left(
    \frac{x'_k-x_k}{R_k}
    \right)_{k\in\mathcal{I}_i}
    \right\|_2.
\end{equation}

The search, therefore, favors successful alarm free candidates that differ as much as possible from the original unstable operating point within the allowed single-participant perturbation region.

%\newpage
\begin{figure*}[tbh]
    \centering
    \includegraphics[width=1.0\columnwidth]{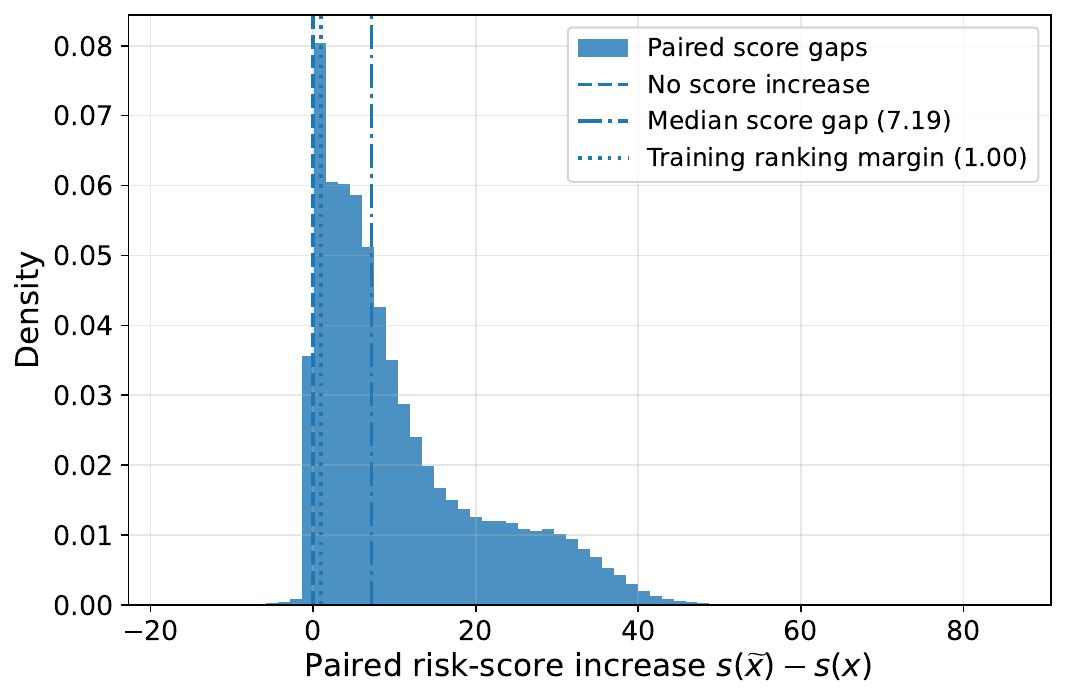}
    \caption{Distribution of the paired risk score increase $s(\widetilde{x})-s(x)$ for pseudo-negatives.}
% generated from held-out stable configurations.}
\label{fig:paired_score_gap}
\end{figure*}

\begin{figure*}[t]
    \centering

    % First row
    \begin{subfigure}[t]{0.48\textwidth}
        \centering
        \includegraphics[
            width=\linewidth
        ]{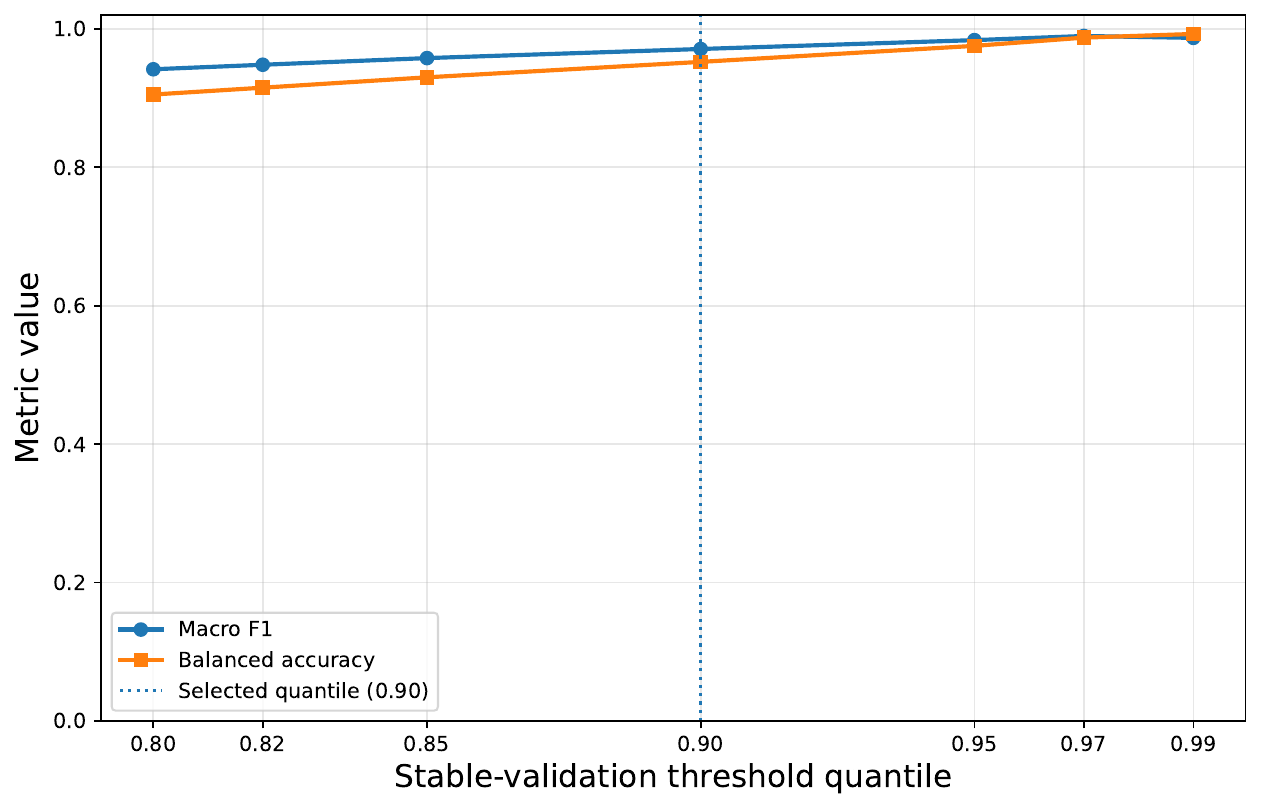}
        \caption{Clean stability prediction.}
        \label{fig:threshold_stability}
    \end{subfigure}
    \hfill
    \begin{subfigure}[t]{0.48\textwidth}
        \centering
        \includegraphics[
            width=\linewidth
        ]{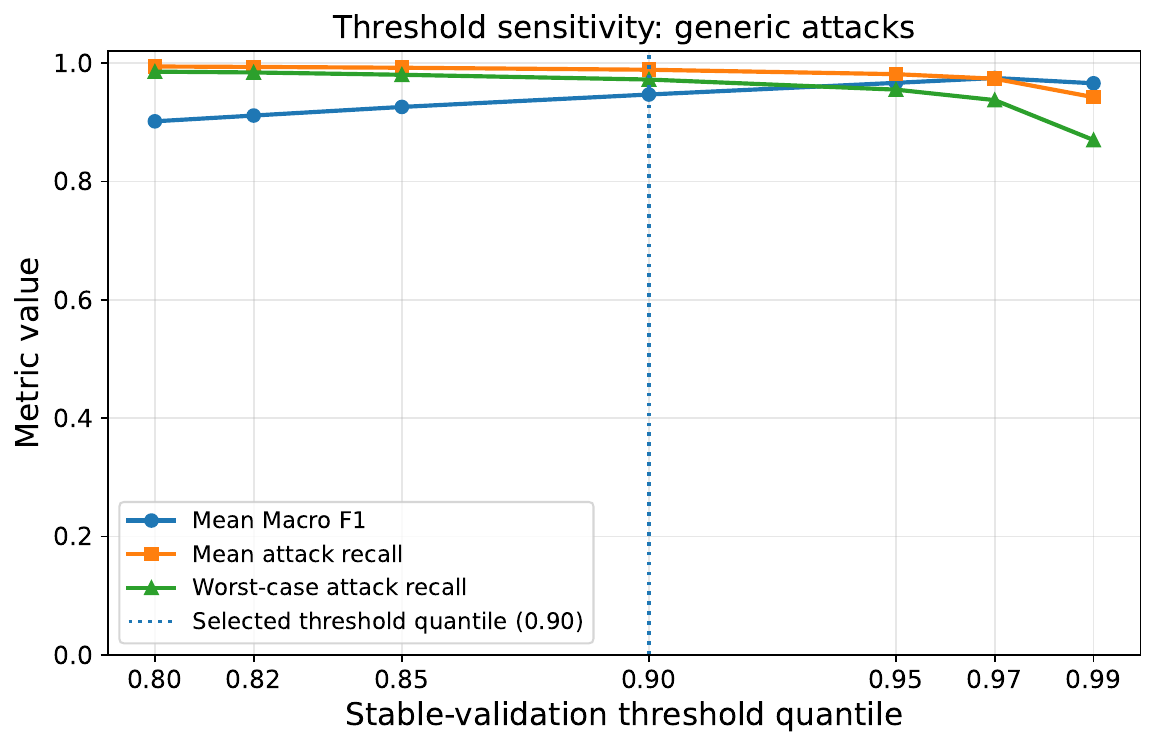}
        \caption{Generic unseen attacks.}
        \label{fig:threshold_generic}
    \end{subfigure}

    \vspace{0.4em}

    % Second row
    \begin{subfigure}[t]{0.48\textwidth}
        \centering
        \includegraphics[
            width=\linewidth
        ]{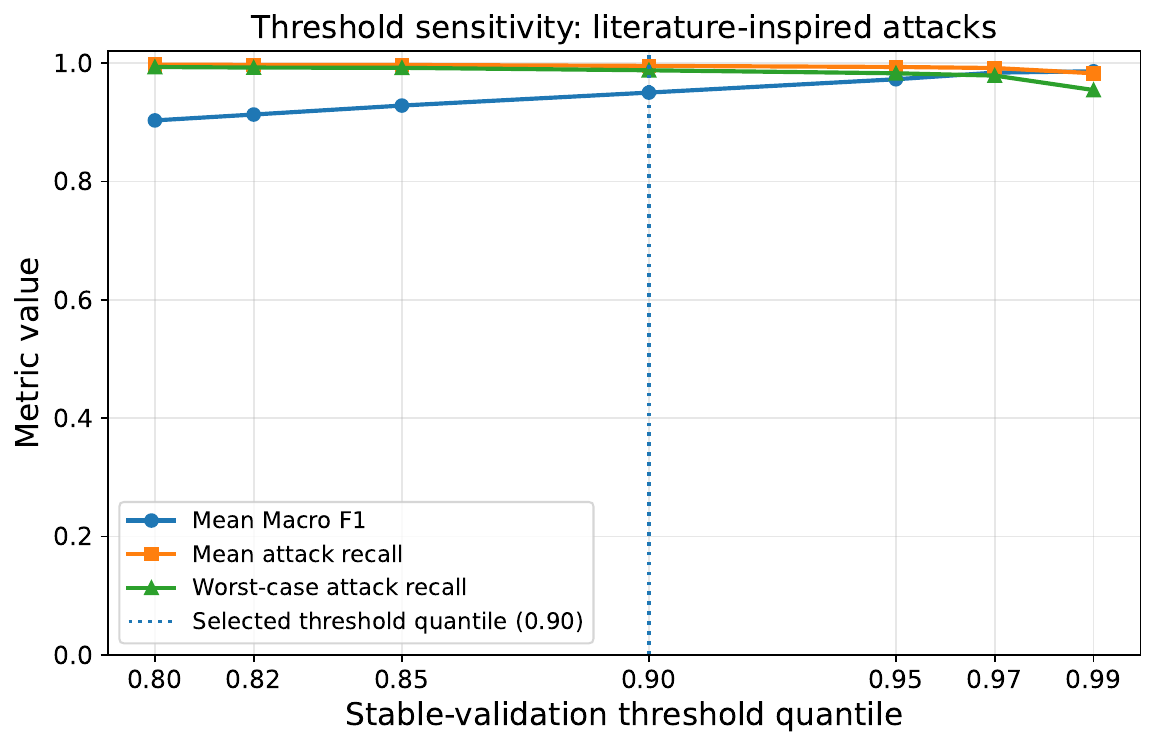}
        \caption{Literature inspired unseen attacks.}
        \label{fig:threshold_literature}
    \end{subfigure}
    \hfill
    \begin{subfigure}[t]{0.48\textwidth}
        \centering
        \includegraphics[
            width=\linewidth
        ]{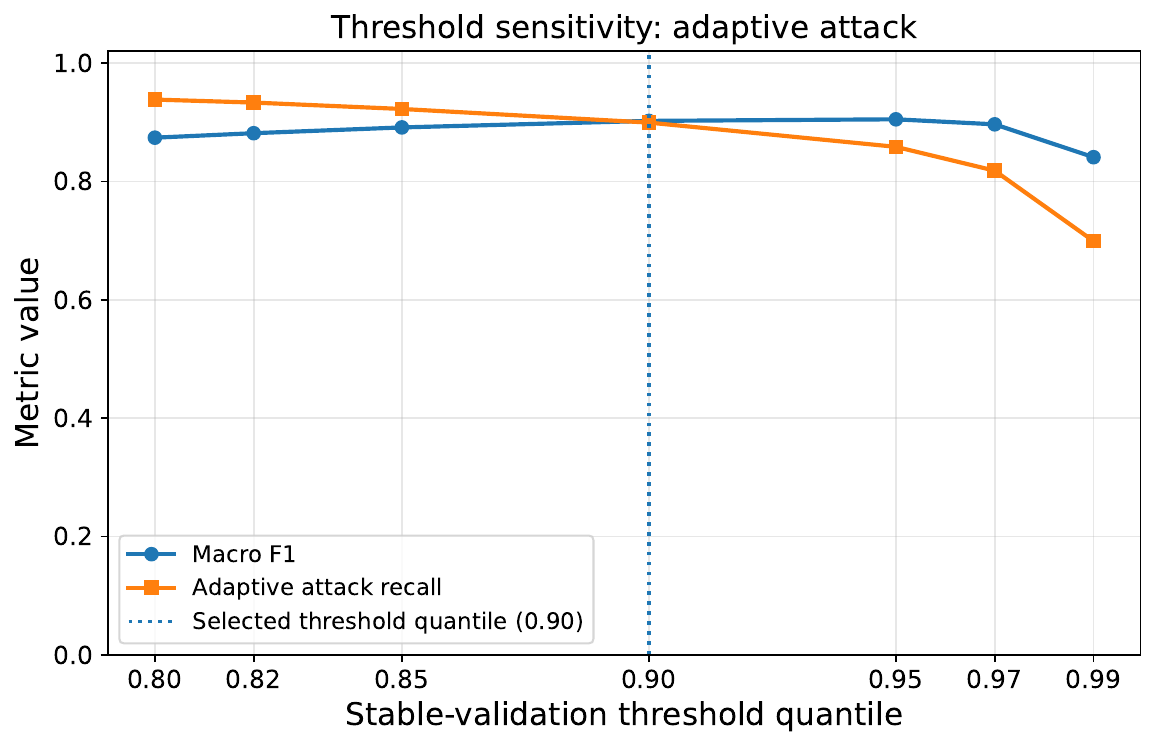}
        \caption{Single node adaptive concealment attack.}
        \label{fig:threshold_adaptive}
    \end{subfigure}

    \caption{
Threshold sensitivity results for clean stability prediction, generic unseen attacks, literature inspired unseen attacks, and the single node adaptive concealment attack. The vertical dotted line denotes the selected operating point, $\alpha=0.90$. The adaptive attack is regenerated independently at each threshold using the corresponding binary detector feedback.
    }
    \label{fig:threshold_ablation}
\end{figure*}

\end{document}